\documentclass[aps,ams,amsmath,prx,longbibliography,twocolumn]{revtex4-1}
\usepackage{graphics}
\usepackage{graphicx}
\usepackage{epstopdf,epsfig}
\usepackage{amsmath}
\usepackage{amssymb}
\usepackage{amsfonts}
\usepackage{mathrsfs}
\usepackage{bm}
\usepackage{xcolor}
\usepackage{bbm}
\usepackage{hyperref}
\usepackage[normalem]{ulem}
\usepackage{comment}
\usepackage{soul}

\newcommand{\vp}{\bm{p}}
\newcommand{\vk}{\bm{k}}
\newcommand{\vq}{\bm{q}}
\newcommand{\vv}{\bm{v}}
\newcommand{\vB}{\bm{B}}
\newcommand{\ep}{\varepsilon}

\begin{document}

\title{Effect of superconducting fluctuations on nonreciprocal dichroism and gyrotropy}

\author{Alex Levchenko}
\affiliation{Department of Physics, University of Wisconsin--Madison, Madison, Wisconsin 53706, USA}

\date{September 25, 2026}

\begin{abstract}
We study the spatially dispersive conductivity of a two-dimensional noncentrosymmetric superconductor, demonstrating that it acquires a nonreciprocal, odd-in-wavevector component from fluctuation-induced Cooper pairs above the critical temperature $T_c$. Utilizing time-dependent Ginzburg-Landau theory generalized to include particle-hole asymmetry and the cubic Lifshitz invariant of trigonal superconductors, we compute the Aslamazov-Larkin contribution to the gyrotropic conductivity in closed form, including its complete frequency dependence. The dissipative part describes nonreciprocal directional dichroism: it is odd in frequency and displays a nonmonotonic dependence, peaking at frequencies comparable to the decay rate of fluctuating Cooper pairs. Its Kramers-Kronig dual component describes gyrotropic birefringence, which remains finite in the static limit and is strongly enhanced as the temperature approaches $T_c$, as long as the probe frequency stays below the pair decay rate. Both effects require simultaneously broken inversion and time-reversal symmetries, are dependent on particle-hole asymmetry in close analogy to the fluctuation Hall effect, and trace to the same asymmetric Cooper-pair dispersion responsible for the superconducting diode effect and the giant magnetochiral anisotropy observed near $T_c$. In this regime this critical enhancement dominates over the smooth normal-state gyrotropy, which we evaluate for the same band model as a baseline. Finally, we frame our analysis within the context of gated transition metal dichalcogenides like MoS$_2$, discussing the implications for probing superconducting dynamics through nitrogen-vacancy-center quantum noise spectroscopy. 
\end{abstract}

\maketitle

\section{Introduction}
\label{sec:intro}

Nonreciprocal transport phenomena, responses that change upon reversal of
the direction of propagation or current flow, have moved to the center of
attention in the physics of quantum materials
\cite{Tokura2018,Ideue2021,Nagaosa2024}. The interest is fueled from several
directions at once: advances in fabrication and exfoliation of
noncentrosymmetric crystals and van der Waals heterostructures, experimental
discoveries of striking transport effects, and a stream of fruitful
theoretical ideas connecting nonreciprocity to band geometry and topology.
Exemplary phenomena include the superconducting diode effect (SDE)
\cite{Ando2020,Daido2022,YuanFu2022,Ilic2022,Nadeem2023,Shaffer2025},
electrical magnetochiral anisotropy (MCA)
\cite{Rikken2001,Rikken1997,Wakatsuki2017,Itahashi2020}, and photogalvanic
phenomena \cite{Kovalev2021,Parafilo2022,Buzdin2024,JTM2026c}.

For the most part these effects require going beyond linear response, which
significantly complicates both theory and the interpretation of experiments.
However, interesting nonreciprocal effects exist already within the domain of
linear response theory. To uncover them one must consider the spatial
dispersion of the conductivity in an external magnetic field. The Onsager
reciprocity principle \cite{Onsager1931} restricts the conductivity tensor to
obey
\begin{equation}
\label{eq:onsager}
\sigma_{ij}(\vk,\omega,\vB)=\sigma_{ji}(-\vk,\omega,-\vB).
\end{equation}
Expanding to leading orders in the wave vector $\vk$ and field $\vB$,
\begin{equation}
\label{eq:expansion}
\sigma_{ij}\simeq\varsigma_{ij}(\omega)+\chi_{ijl}(\omega)k_l
+\lambda_{ijl}(\omega)B_l+g_{ijlm}(\omega)\,k_l B_m ,
\end{equation}
one identifies, besides the Drude part $\varsigma_{ij}$, the Hall response $\lambda_{ijl}$, the natural optical activity $\chi_{ijl}$, and the gyrotropic birefringence $g_{ijlm}$ \cite{LandauLifshitzECM}, 
which describes nonreciprocal propagation of electromagnetic waves in a medium \cite{Hornreich1968}. By Eq.~\eqref{eq:onsager}, $\chi$ and $\lambda$ are antisymmetric in $(ij)$
while $\varsigma$ and $g$ are symmetric; $\chi$ requires broken inversion symmetry, $\lambda$ broken time-reversal symmetry, and the gyrotropic tensor $g$ requires both to be broken. 
It has been recognized that these tensors are sensitive probes of band geometry: natural optical activity is governed by the orbital magnetic moment of quasiparticles
\cite{Ma2015,Zhong2016,Shinada2024}, and the anomalous Hall response by the Berry curvature \cite{Nagaosa2010AHE,Niu2010}.

Beyond the single-particle level, including the well-developed topological
band theory of these coefficients, comparatively little is known about the
impact of electron interactions on this intricate family of effects. One
recent lesson is that the magnetochiral anisotropy is strongly enhanced by
the proliferation of superconducting correlations upon approaching the
critical temperature $T_c$, as observed in gated MoS$_2$
\cite{Wakatsuki2017} and in polar SrTiO$_3$ \cite{Itahashi2020}. These
observations triggered multiple theoretical studies of the fluctuation MCA
\cite{Hoshino2018,WakatsukiNagaosa2018,JTM2026a,JTM2026b}, nonlinear Hall effect \cite{Daido2024,Boeva2024,Dong2025}, 
and of photogalvanic effects \cite{Kovalev2021,Parafilo2022,JTM2026c} arising from fluctuations. 
The common microscopic origin of the enhancement was identified in the Lifshitz invariants (LI) of the
Ginzburg-Landau (GL) free energy: terms odd in the Cooper-pair momentum
$\vq$ allowed by broken inversion symmetry (other possible mechanisms arising from the parity mixing and Berry dipole could also be at play). 
The invariants linear in $\vq$ have been systematically tabulated for all crystallographic point groups
and derived microscopically for several band models \cite{MineevSamokhin1994,Kaur2005,Agterberg2012,Smidman2017}. 
Higher-order (cubic) LI have been intensively studied recently, mostly for noncentrosymmetric superconductors with Rashba- or Ising-type spin-orbit
coupling, where they control the diode effect and the MCA \cite{Daido2022,YuanFu2022,HeLaw2022,Ilic2022,Shaffer2024}.

In this work we build on these advances and extend the theory of superconducting fluctuations to the linear-response nonreciprocal coefficients:
the dichroism and gyrotropy near $T_c$ of superconducting
transition metal dichalcogenides, with the band model of MoS$_2$
\cite{Lu2015,Saito2016,Wakatsuki2017} as a guiding example. 
The main findings of this work can be summarized as follows.  
(i) A purely linear LI produces no nonreciprocal
conductivity: it can be removed by a shift of the pair momentum and only
renormalizes $T_c(\vB)$, the transport counterpart of the known
argument \cite{Agterberg2012} that forbids SDE in the helical phase to this order. 
The effect is controlled by the next order cubic LI. (ii) Even the cubic LI is silent in the standard
time-dependent GL (TDGL) description: the fluctuation gyrotropy
also requires particle-hole asymmetry, entering through the imaginary part
of the TDGL relaxation constant, in precise analogy with the fluctuation Hall
and anomalous Nernst effects
\cite{FukuyamaEbisawaTsuzuki1971,AronovHikamiLarkin1995,LiLevchenko2020}.
(iii) The resulting response has a nonmonotonic frequency dependence
that we obtain in closed form: the dissipative (dichroic) part is odd in
$\omega$, vanishes at dc, and peaks at $\omega\tau_{\rm GL}\simeq2$, where $\tau_{\rm GL}$ is the GL relaxation time, while
the reactive (birefringent) part survives at $\omega\to0$ and diverges as
$1/(T-T_c)$ leading to a critical enhancement of gyrotropy; at a fixed
probe frequency this enhancement is cut off once $\omega\tau_{\rm GL}\gtrsim1$,
i.e., within $T-T_c\simeq\pi\hbar\omega/16k_B$ of the transition.

\section{Technical framework}
\label{sec:framework}

Transport anomalies above $T_c$ are governed by three fluctuation
mechanisms: the Aslamazov-Larkin (AL) paraconductivity
\cite{AslamazovLarkin1968}, the Maki-Thompson interference contribution
\cite{Maki1968,Thompson1970}, and the density-of-states suppression
\cite{Abrahams1970}; together they account for the observed fluctuation-driven
anomalies in transport and thermodynamics of superconductors
\cite{VarlamovLarkin}. In systems with pair breaking, to which a magnetic field contributes significantly, the
AL process  is usually the leading mechanism. 
Early on, Schmid \cite{Schmid1969} showed that an efficient way to
capture it is through the TDGL equation, 
which was derived by Gor'kov and Eliashberg for the gapless state of a superconductor \cite{GE1968}, 
supplemented by Langevin forces
describing thermally excited fluctuations of the order parameter. We briefly
recapitulate this framework with the generalization needed here: the account
of particle-hole asymmetry and of the Lifshitz invariants.

Consider the Gaussian GL free energy of a two-dimensional superconductor
without an inversion center,
\begin{equation}
\label{eq:GL}
F[\Psi]=\sum_{\vq}\ep_{\vq}\,|\Psi_{\vq}|^{2},\qquad
\ep_{\vq}=a+\frac{q^{2}}{2m}+w(\vq),
\end{equation}
where $a=\alpha\epsilon$ with $\epsilon=\ln(T/T_c)\simeq (T-T_c)/T_c$, $m$
is the effective pair mass, and $w(-\vq)=-w(\vq)$ collects the
inversion-odd terms. Time-reversal symmetry forces the coefficients of
$w$ to be odd in $\vB$. For the trigonal point group $D_{3h}$, relevant to
monolayer and gated few-layer MoS$_2$ with an out-of-plane field, symmetry
allows no linear invariant and the leading term is cubic
\cite{Wakatsuki2017},
\begin{equation}
\label{eq:LI-D3h}
w_{D_{3h}}(\vq)=\eta B_z\,(q_x^{3}-3q_xq_y^{2}),
\end{equation}
while for the Rashba point group $C_{3v}$ (polar axis $\hat z$, in-plane
field) the linear invariant $\propto[\hat{\bm{z}}\times\vB]\cdot\vq$ coexists with
cubic ones such as \cite{JTM2026b}
\begin{equation}
\label{eq:LI-Rashba}
w_{C_{3v}}(\vq)=\eta\, (q_x^{2}+q_y^{2})(q_xB_y-q_yB_x).
\end{equation}
Microscopically $\eta B_z\propto \lambda\,\Delta_{\rm SO}\,\Delta_Z/T_c^{2}$
for the $D_{3h}$ case, where $\lambda$ is the trigonal warping of the bands,
$\Delta_{\rm SO}$ the Ising spin-orbit splitting, and
$\Delta_Z=\frac12 g\mu_B B_z$ the Zeeman energy
(see Appendix~\ref{app:normal}); the valley orbital magnetic moment
$m_vB_z$, which carries the same valley-odd, spin-even, time-reversal-odd
quantum numbers as the product $(vs\Delta_{\rm SO})(s\Delta_Z)$, contributes
a term of the same symmetry to $\eta$, at order $\lambda m_vB_z/T_c$, so that
$\eta$ is treated below as a phenomenological coefficient.

The relaxational dynamics of the fluctuating order parameter is
\begin{equation}
\label{eq:TDGL}
\gamma\,\partial_t\Psi=-\frac{\delta F}{\delta\Psi^{*}}+\zeta
=-\Big[a-\frac{\nabla^{2}}{2m}+w(-i\nabla)\Big]\Psi+\zeta ,
\end{equation}
with a complex relaxation constant
\begin{equation}
\gamma=\gamma_1+i\gamma_2,\qquad
\frac{\gamma_2}{\gamma_1}\propto\frac{\partial\ln T_c}{\partial\ln\mu},
\end{equation}
whose imaginary part encodes the particle-hole asymmetry of the pairing
interaction \cite{FukuyamaEbisawaTsuzuki1971,AronovHikamiLarkin1995};
up to a model-dependent factor of order unity,
$\gamma_2/\gamma_1\sim(T_c/E_F)\,\partial\ln T_c/\partial\ln E_F$, so that its
sign is set by the slope of $T_c$ versus carrier density, positive on the
underdoped and negative on the overdoped side of a superconducting dome, and
its magnitude is of order $10^{-2}$ for gated MoS$_2$.
Writing Eq.~\eqref{eq:TDGL} as
$\partial_t\Psi=-\gamma^{-1}\delta F/\delta\Psi^{*}+\gamma^{-1}\zeta$, the
mobility is $\gamma^{-1}$ and the stationary measure is the Gibbs weight
$e^{-F/T}$ provided the Langevin force obeys the fluctuation-dissipation
relation
\begin{equation}
\label{eq:FDT}
\langle\zeta(\vq,\nu)\zeta^{*}(\vq',\nu')\rangle
=2\gamma_1 T\,(2\pi)^{3}\delta(\vq-\vq')\delta(\nu-\nu').
\end{equation}
Only the dissipative $\gamma_1$ enters the noise; the reactive
$i\gamma_2\partial_t$ does no work, and the equal-time (Gibbs) fluctuations
$\langle|\Psi_{\vq}|^2\rangle=T/\ep_{\vq}$ are independent of $\gamma_2$.
Particle-hole asymmetry thus resides entirely in the dynamics. Solving
Eq.~\eqref{eq:TDGL}, $\Psi(\vq,\nu)=\zeta(\vq,\nu)/(\ep_{\vq}-i\gamma\nu)$, one
finds the statistical pair propagator
\begin{equation}
\label{eq:propagator}
\Pi(\vq,\nu)=\langle|\Psi(\vq,\nu)|^{2}\rangle
=\frac{2\gamma_1 T}{(\ep_{\vq}+\gamma_2\nu)^{2}+\gamma_1^{2}\nu^{2}} ,
\end{equation}
which is not even in $\nu$ once $\gamma_2\neq0$, the property that
ultimately unlocks the nonreciprocal response.
The natural question arises as to how nonreciprocal terms modify the noise correlation function, which was not included in Eq. \eqref{eq:FDT}. 
This question is addressed in Appendix \ref{app:noise}, where we provide a detailed explanation showing that these terms do not modify the central results concerning gyrotropy to leading order near $T_c$.

Gauge invariance ($\vq\to\vq-2e\bm{A}$ for a pair of charge $2e$) fixes the
current vertex to the pair group velocity,
\begin{equation}
\label{eq:vertex}
\bm{J}=2e\,\vv(\vq)\,,\qquad
\vv(\vq)=\frac{\partial\ep_{\vq}}{\partial\vq}
=\frac{\vq}{m}+\frac{\partial w}{\partial\vq} .
\end{equation}
The inversion-odd anomalous velocity $\partial w/\partial\vq$ is an
essential part of the vertex; as shown below it in fact dominates the
nonreciprocal response. Evaluating the classical Kubo formula for the
symmetrized current-current correlator with Wick contractions of the
Gaussian field $\Psi$, the dissipative part of the fluctuation conductivity
collapses to a single loop,
\begin{align}
\label{eq:bubble}
\mathrm{Re}\,\sigma_{ij}(\vk,\omega)
&=\frac{(2e)^{2}}{2T}\!\int\!\frac{d^{2}q\,d\nu}{(2\pi)^{3}}\;
v_i(\vq)\,v_j(\vq)\nonumber\\
&\quad\times\Pi\big(\vq_{+},\nu_{+}\big)\,\Pi\big(\vq_{-},\nu_{-}\big),
\end{align}
with $\vq_\pm=\vq\pm\vk/2$ and $\nu_\pm=\nu\pm\omega/2$; here $\omega$ and
$\vk$ are the frequency and wave vector of the probing field. Being a
symmetrized correlator, Eq.~\eqref{eq:bubble} is the dissipative
(absorptive) part of the response; the reactive part follows from the
Kramers-Kronig relation. Equation~\eqref{eq:bubble} is the central object
of this paper: all reciprocal and nonreciprocal fluctuation transport
discussed below is contained in the symmetry structure of the two-propagator
kernel \footnote{Gauge invariance fixes the exact paramagnetic
vertex $\Gamma_i(\vq_+,\vq_-)$ through the Ward identity
$k_i\Gamma_i(\vq_+,\vq_-)=\ep_{\vq_+}-\ep_{\vq_-}$ (for a polynomial dispersion
$\Gamma_i$ is the corresponding divided difference); it is even in $\vk$ to
all orders and reduces to $v_i(\vq)$ at linear order, so all odd-in-$\vk$
dependence of Eq.~\eqref{eq:bubble} resides in the propagators. The
diamagnetic term is even in $\vk$ as well, being symmetric under
$\bm{A}_{\vk}\bm{A}_{-\vk}$, and contributes only to the reactive part of the
reciprocal sector.}. In the following, we split the dispersive conductivity into even and odd parts, 
\begin{equation}
\mathrm{Re}\,\sigma_{ij}(\vk,\omega)=\mathrm{Re}\,\sigma^{\text{even}}_{ij}(\vk,\omega)+\mathrm{Re}\,\sigma^{\text{odd}}_{ij}(\vk,\omega),
\end{equation}
and discuss each term separately.

We close this section by stating the range of validity of the framework.
The Gaussian TDGL description requires ${\rm Gi}\ll\epsilon\ll1$, where the
Ginzburg number is ${\rm Gi}\sim T_c/E_F\sim10^{-2}$ for clean gated MoS$_2$;
classical pair dynamics, $\hbar\omega\ll k_BT$, which in terms of the variables
used below reads $\hbar\omega/k_BT=(16/\pi)\epsilon\,\omega\tau_{\rm GL}$; and
long wavelengths, $k\xi\ll1$. Above $T_c$ the symmetry of the condensate plays
no role: whatever the pairing channel, only the softest pair mode is critical
near $T_c$, and it is described by a single complex field $\Psi$, which
justifies the scalar order parameter for multiband materials such as
NbSe$_2$ as well as for MoS$_2$. The magnetic field enters through the
Zeeman-driven Lifshitz invariant, with $\Delta_Z\ll T_c$; its orbital
coupling to the pair motion is even in $B$ and is neglected, which requires
$B\ll H_{c2}(T)$, i.e., $\epsilon\gg\epsilon_B\equiv B/T_c|dH_{c2}/dT|$. Pair
breaking of rate $\delta$ (orbital, Zeeman, or spin-flip) shifts
$\epsilon\to\epsilon+\delta$ in all formulas below without altering the
frequency functions, and disorder enters only through the phenomenological
coefficients $m$, $\gamma$, and $\eta$. Finally, only the AL channel is
retained: the Maki-Thompson and density-of-states contributions are subject
to the same symmetry constraints (particle-hole asymmetry and odd velocity
content), the former is suppressed by pair breaking and has a different vertex
structure with its own point-group selection rules, and the latter is less
singular in $\epsilon$; their nonreciprocal parts are left for future work.

\section{Reciprocal part: dispersive paraconductivity}
\label{sec:even}

Consider first $\gamma_2=0$. The internal frequency integral in
Eq.~\eqref{eq:bubble} is elementary,
\begin{equation}
\label{eq:K0}
\int\!\frac{d\nu}{2\pi}\,\Pi(\ep_+,\nu_+)\Pi(\ep_-,\nu_-)
=\frac{2\gamma_1T^{2}(\ep_++\ep_-)}
{\ep_+\ep_-\big[(\ep_++\ep_-)^{2}+\gamma_1^{2}\omega^{2}\big]},
\end{equation}
with $\ep_\pm\equiv\ep_{\vq\pm\vk/2}$. This kernel is a symmetric
function of $\ep_+$ and $\ep_-$; since $\vk\to-\vk$ merely exchanges them,
the conductivity is strictly even in $\vk$ for real $\gamma$ for an
arbitrary inversion-odd $w(\vq)$, cubic terms included. For a purely linear
LI, $w=\bm{c}\cdot\vq$ with $\bm{c}\propto\vB$, the shift
$\vq\to\vq-m\bm{c}$ maps the problem exactly onto the centrosymmetric one
with $a\to a-mc^{2}/2$: the linear invariant only shifts $T_c(\vB)$
\cite{Agterberg2012} and drops out of the nonlocal response entirely.

The even-in-$\vk$ conductivity is the dispersive generalization of the AL
paraconductivity. Performing the remaining momentum integrals in two
dimensions at $\omega\to0$ we obtain
\begin{equation}
\label{eq:FTL}
\sigma^{\text{even}}_{ij}(\vk)=\sigma_{\rm AL}
\Big[\Big(\delta_{ij}-\hat k_i\hat k_j\Big)F_T(\kappa)
+\hat k_i\hat k_j F_L(\kappa)\Big],
\end{equation}
with $\kappa=k\xi/2$, $\xi^{2}=1/2ma$ the GL coherence length,
$\sigma_{\rm AL}=\gamma_1 T (2e)^2/8\pi a$ the AL conductivity (equal to the
celebrated $e^{2}/16\hbar\epsilon$ for the microscopic value of $\gamma_1$
\cite{AslamazovLarkin1968,Schmid1969,VarlamovLarkin}), and the scaling
functions
\begin{subequations}
\begin{align}
\label{eq:FL2D}
F_L(\kappa)&=\frac{\ln(1+\kappa^{2})}{\kappa^{2}},\\
\label{eq:FT2D}
F_T(\kappa)&=\frac{2\,\mathrm{arcsinh}\,\kappa}{\kappa\sqrt{1+\kappa^{2}}}
-\frac{\ln(1+\kappa^{2})}{\kappa^{2}} ,
\end{align}
\end{subequations}
which interpolate between $F_{T,L}\to1$ at $\kappa\to0$
[$F_T\simeq1-\tfrac56\kappa^{2}$, $F_L\simeq1-\tfrac12\kappa^{2}$] and a
slow decay at $\kappa\gg1$; $F_L$ is the bare-bubble longitudinal
function, which in the static limit at finite $\vk$ is subject to the
charge-conservation (screening) corrections discussed in
Appendix~\ref{app:normal}, whereas the transverse function $F_T$ is not. At finite frequency the same kernel yields
$F_{T,L}(\kappa,\varpi)$ depending additionally on
$\varpi=\omega\tau_{\rm GL}$ (with $\tau_{\text{GL}}=\gamma_1/2a$ the GL relaxation time). 
We note that for $\bm{k}\to0$ optical paraconductivity was calculated in Ref. \cite{AslamazovVarlamov1980}.
The dispersive fluctuation conductivity at $\omega\to0$ was derived in Refs. \cite{Barash1993,Galaktionov1995}. 

Beyond completeness, Eqs.~\eqref{eq:FTL}-\eqref{eq:FT2D} have a direct
experimental application: the wave-vector-resolved dissipative conductivity
$\mathrm{Re}\,\sigma(\vk,\omega)$ determines the magnetic noise sensed by a
proximate spin qubit, and thereby the relaxation ($1/T_1$) and decoherence
rates measured in nitrogen-vacancy magnetometry
\cite{Dolgirev2022,Chatterjee2022}. Recent quantum noise spectroscopy of
superconducting critical dynamics \cite{Liu2025} in a BSCCO thin film probes precisely this
object at $k\sim1/z_{\rm NV}$ set by the qubit-sample distance, making the
dispersive paraconductivity \eqref{eq:FTL} directly measurable. 
The reported noise enhancement near $T_c$ is consistent with 
$1/T_1\propto T\mathrm{Re}\,\sigma(k,\omega_{\text{NV}})\propto \sigma_{\rm AL}$ expected
in the Gaussian regime.

\section{Nonreciprocal part: fluctuation dichroism and gyrotropy}
\label{sec:odd}

We now restore $\gamma_2$ and extract the odd-in-$\vk$ response. Expanding
the frequency-integrated kernel $K=\int\frac{d\nu}{2\pi}\Pi\Pi$ in Eq. \eqref{eq:bubble} to linear order in $\gamma_2$ and in
$\delta\varepsilon\equiv\ep_+-\ep_-=\vk\cdot\vv(\vq)+O(k^{3})$, we find the compact result
\begin{equation}
\label{eq:deltaK}
\delta K=-8\gamma_1\gamma_2T^{2}
\frac{(\vk\cdot\vv)\omega}
{\ep\,\big(4\ep^{2}+\gamma_1^{2}\omega^{2}\big)^{2}},
\end{equation}
with $\ep\equiv\ep(\vq)$. The structure of Eq.~\eqref{eq:deltaK} makes the
selection rules explicit: the odd response is proportional to the
product of the particle-hole asymmetry $\gamma_2$ and the
inversion-odd velocity content of $\vk\cdot\vv$, and it is odd in the
external frequency $\omega$. Either factor alone gives zero: without
$\gamma_2$ the kernel \eqref{eq:K0} is symmetric, and without the
(irremovable, cubic) LI the angular average of $\vk\cdot\vv$ carries no odd
harmonics. Substituting into Eq.~\eqref{eq:bubble} gives the main formula
\begin{equation}
\label{eq:master}
\mathrm{Re}\,\sigma^{\rm odd}_{ij}(\vk,\omega)
=-4(2e)^{2}T\gamma_1\gamma_2\,\omega\!\int\!\frac{d^{2}q}{(2\pi)^{2}}
\frac{v_i v_j\,(\vk\cdot\vv)}
{\ep\,\big(4\ep^{2}+\gamma_1^{2}\omega^{2}\big)^{2}},
\end{equation}
obtained to linear order in $\gamma_2$ and $\vk$, with the full dependence on
$\omega$ retained.

Evaluating Eq.~\eqref{eq:master} with the cubic invariant
\eqref{eq:LI-D3h} to linear order in $\eta B_z$, both the anomalous-vertex
channel [$\partial w/\partial\vq$ in $\vv$] and the propagator channel [$w(\vq)$
in $\ep$] contribute, in the ratio $3:(-1)$ at all frequencies (see below); their sum gives, for
$\vk\parallel\hat x$,
\begin{equation}
\label{eq:mainresult}
\mathrm{Re}\,\sigma^{\rm odd}_{xx}(\vk,\omega)
=-\frac{(2e)^{2}T\,\gamma_2\,m\,\eta B_z}{8\pi a}\;k_x\,
\mathcal{W}(\varpi)
\end{equation}
with the dimensionless frequency $\varpi=\omega\tau_{\rm GL}$,
$\tau_{\rm GL}=\gamma_1/2a$, and the closed-form frequency function
\begin{equation}
\label{eq:W}
\mathcal{W}(\varpi)=\frac{6}{\varpi}
-\frac{12\arctan\varpi}{\varpi^{2}}
+\frac{6\ln(1+\varpi^{2})}{\varpi^{3}} .
\end{equation}
The full tensor is symmetric and traceless with trigonal structure,
\begin{equation}
\label{eq:tensor}
\big(\sigma^{\rm odd}_{xx}-\sigma^{\rm odd}_{yy}\big)
+2i\,\sigma^{\rm odd}_{xy}\;\propto\;\eta B_z\,(k_x-ik_y),
\end{equation}
i.e., the $\ell=2$ harmonic obtained by combining the $\ell=3$ warping with
one power of $\vk$, as dictated by the threefold axis. This is the
symmetric gyrotropic tensor $g_{ijlm}k_lB_m$ of Eq.~\eqref{eq:expansion};
the antisymmetric natural-activity tensor $\chi$ is not generated by
the scalar AL bubble.

The ratio $3{:}(-1)$ of the two channels is independent of frequency and
of the form of the kernel. Writing the integrand of Eq.~\eqref{eq:master} as
$v_iv_j\,\partial_l\Phi(\ep)$ with $\Phi'(\ep)=[\ep(4\ep^2+\gamma_1^2\omega^2)^2]^{-1}$
and integrating by parts, the response at linear order in $w$ collapses to
the single integral $-(2/m)\int_{\vq}\Phi(\ep^{(0)}_{\vq})\,q_{(i}\partial_{j)}\partial_lw$,
with $\ep^{(0)}_{\vq}=a+q^2/2m$ and $\int_{\vq}\equiv\int d^2q/(2\pi)^2$ [the remaining term, $\propto\int_{\vq} w\,\Phi'\,q_j$,
vanishes for an $\ell=3$ harmonic]. The vertex and propagator pieces of this
integral are fixed at $3{:}(-1)$ by the angular averages
$\langle\cos^2\theta\cos2\theta\rangle=1/4$ and
$\langle\cos^3\theta\cos3\theta\rangle=1/8$ together with one radial
integration by parts, for any weight function $\Phi$. The same identity
applies to the third velocity moment of the normal state,
Eq.~\eqref{eq:NSM}, which is why the ratio recurs in Appendix~\ref{app:normal}:
it is a property of the parabolic band and of the $\ell=3$ harmonic, not of
the kernel.

\begin{figure}[t]
\centering
\includegraphics[width=\columnwidth]{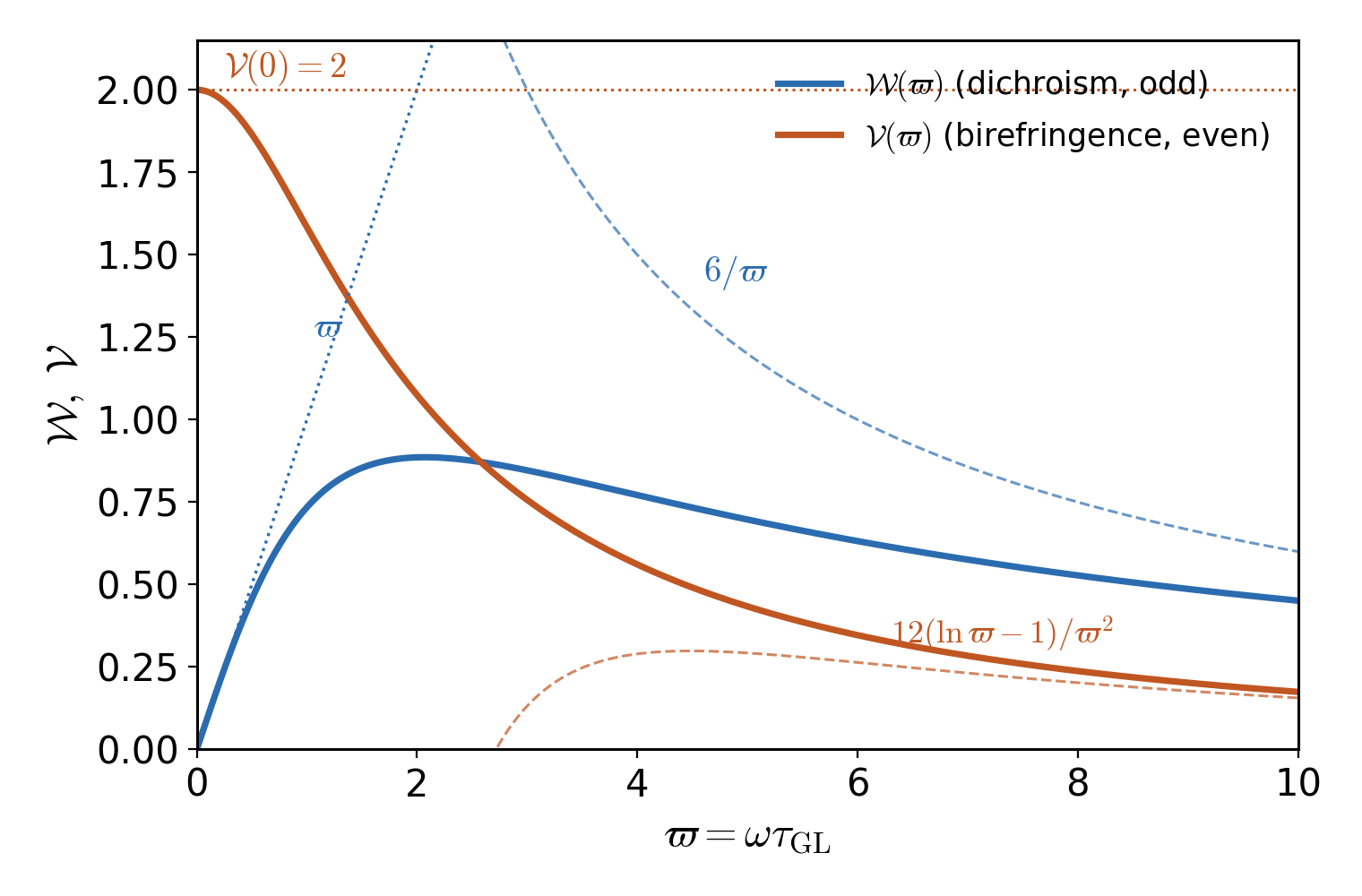}
\caption{Frequency functions of the nonreciprocal fluctuation conductivity
versus $\varpi=\omega\tau_{\rm GL}$. The dissipative function
$\mathcal{W}(\varpi)$ [Eq.~\eqref{eq:W}, blue] controls nonreciprocal
directional dichroism; it is odd in $\omega$, rises as $\varpi$, peaks at
$\varpi\simeq2.07$, and decays as $6/\varpi$. Its Kramers--Kronig partner
$\mathcal{V}(\varpi)$ [Eq.~\eqref{eq:V}, orange] controls gyrotropic
birefringence; it is even in $\omega$, with $\mathcal{V}(0)=2$ and the tail
$12(\ln\varpi-1)/\varpi^{2}$. Dotted and dashed lines show the asymptotes.}
\label{fig:freq}
\end{figure}

The function $\mathcal{W}$ is manifestly odd in $\omega$, with the limits
\begin{equation}
\label{eq:Wlimits}
\mathcal{W}(\varpi\to0)=\varpi-\tfrac{2}{5}\varpi^{3}+\dots,
\qquad
\mathcal{W}(\varpi\to\infty)=\frac{6}{\varpi},
\end{equation}
and a maximum $\mathcal{W}\simeq0.886$ at $\varpi\simeq2.07$
(Fig.~\ref{fig:freq}). In the hydrodynamic regime
$\omega\tau_{\rm GL}\ll1$ the dichroic response grows as
$\mathrm{Re}\,\sigma^{\rm odd}\propto\eta B_z\,\omega k/\epsilon^{2}$, 
one power of $\epsilon=(T-T_c)/T_c$ more singular than the AL
paraconductivity.

Because the $\vk$-odd, $\vB$-linear coefficient
$g(\omega)$ is a causal response function, its real and imaginary parts are
Hilbert conjugates. Both descend from a single function analytic in the
upper half-plane of $\varpi$, in terms of which the full complex
nonreciprocal conductivity reads
$\sigma^{\rm odd}_{xx}(\vk,\omega)=i\,[(2e)^{2}T\gamma_2m\eta B_z/8\pi a]\,k_x\,\mathcal{G}(\varpi)$,
\begin{align}
\label{eq:G}
\mathcal{G}(\varpi)&=\mathcal{V}(\varpi)+i\mathcal{W}(\varpi)
=6\!\int_{1}^{\infty}\!\frac{(u-1)^{2}}{u^{2}(u-i\varpi)^{2}}\,du
\nonumber\\
&=\frac{12i(1-i\varpi)\ln(1-i\varpi)}{\varpi^{3}}
-\frac{6(2-i\varpi)}{\varpi^{2}} ,
\end{align}
whose imaginary part reproduces Eq.~\eqref{eq:W} and whose real part,
\begin{equation}
\label{eq:V}
\mathcal{V}(\varpi)
=6\!\int_{1}^{\infty}\!\frac{(u-1)^{2}(u^{2}-\varpi^{2})}
{u^{2}(u^{2}+\varpi^{2})^{2}}\,du ,
\end{equation}
is even in $\omega$ with the exact limits
\begin{equation}
\label{eq:Vlimits}
\mathcal{V}(0)=2,\qquad
\mathcal{V}(\varpi\to\infty)=\frac{12(\ln\varpi-1)}{\varpi^{2}} .
\end{equation}
The reactive nonreciprocal response therefore survives in the static
limit: fluctuations produce a temperature-dependent gyrotropic
birefringence
\begin{equation}
\label{eq:staticbiref}
\mathrm{Im}\,\sigma^{\rm odd}_{xx}(\vk,\omega\to0)
=\frac{(2e)^{2}T\,\gamma_2\,m\,\eta B_z}{4\pi a}\,k_x
\;\propto\;\frac{k_x B_z}{\epsilon},
\end{equation}
enhanced as $1/\epsilon$ on approach to $T_c$ at fixed $\varpi$, while the dissipative
dichroism switches on linearly in $\omega$ and dominates for
$\omega\tau_{\rm GL}\gtrsim1$. Both functions are shown in
Fig.~\ref{fig:freq}.

Three remarks are in order concerning the limits involved. First,
Eqs.~\eqref{eq:deltaK}--\eqref{eq:staticbiref} are the coefficient of the
term linear in $\vk$ at fixed $\omega$, valid for $k\xi\ll1$, and the limit
$\omega\to0$ is taken last. Since the coefficient $g(\omega)$ is regular at
$\omega=0$, the two limits commute and Eq.~\eqref{eq:staticbiref} is a
static, reactive spatial-dispersion response: in real space it is the
nondissipative current $j_x=g''\,\partial_xE_x$ driven by the gradient of
the field, rather than a dc conductivity. In the longitudinal channel the
static limit at finite $\vk$ is further subject to charge conservation,
which screens the longitudinal response through the diffusion pole discussed
in Appendix~\ref{app:normal}; this does not affect the transverse channel or
propagating-wave probes, for which $Dk^2\ll\omega$. Second, the $1/\epsilon$
laws above are statements at fixed $\varpi$. In an experiment at fixed probe
frequency, $\varpi=\pi\hbar\omega/16k_B(T-T_c)$ grows on approach to $T_c$
(for the BCS value $\gamma_1=\pi\alpha/8T$), and since the prefactor of
Eq.~\eqref{eq:mainresult} is itself $\propto\varpi/\omega$, the temperature
dependence is governed by $\varpi\mathcal{W}(\varpi)$ and
$\varpi\mathcal{V}(\varpi)$: the dichroism grows as $1/\epsilon^2$ and
saturates at $|\mathrm{Re}\,\sigma^{\rm odd}_{xx}|=(3/2\pi)(2e)^2T(\gamma_2/\gamma_1)m\eta B_zk_x/\omega$
once $\omega\tau_{\rm GL}\gg1$, while the birefringence grows as $1/\epsilon$,
peaks at $\omega\tau_{\rm GL}\simeq3.2$ with the value
$\simeq2.3\,(2e)^2T(\gamma_2/\gamma_1)m\eta B_zk_x/4\pi\omega$, and then
decreases as $\epsilon\ln(1/\epsilon)$. The crossover
$T^*-T_c=\pi\hbar\omega/16k_B\approx9\,{\rm mK}\times(\omega/2\pi)/{\rm GHz}$
lies below the Ginzburg region for microwave probes, so that the $1/\epsilon$
law of Eq.~\eqref{eq:staticbiref} applies throughout the Gaussian regime
there, whereas THz probes operate in the saturated regime; in both cases the
maximal attainable response scales as $1/\omega$. Third, relative to the
paraconductivity the effect is
$\mathrm{Im}\,\sigma^{\rm odd}_{xx}(\vk,0)/\sigma_{\rm AL}=2(\gamma_2/\gamma_1)\,m\eta B_zk_x=(\gamma_2/\gamma_1)\,\delta_{\rm LI}\,k_x\xi_0$,
where $\delta_{\rm LI}\equiv2m\eta B_z/\xi_0$ measures the Lifshitz term
against the pair kinetic energy at the coherence scale $\xi_0$; with
$\gamma_2/\gamma_1\sim10^{-2}$ this is $\sim10^{-2}\delta_{\rm LI}\,k\xi_0$,
which favors near-field probes ($k\xi_0\sim0.1$--$0.3$) over far-field
optics ($k\xi_0\sim10^{-4}$).

\section{Discussion and Outlook}
\label{sec:discussion}

The dependence of the effect on $\gamma_2$ is not an artifact of the TDGL
approximation but has the same origin as the well-known suppression of the
fluctuation Hall and anomalous Nernst responses at particle-hole symmetry
\cite{FukuyamaEbisawaTsuzuki1971,AronovHikamiLarkin1995}.
With real $\gamma$ the AL bubble is built from the squared
modulus of the pair propagator, a symmetric function of the two pair
energies. Such a bubble supports only dissipative, reciprocal, longitudinal
response. Any reactive or nonreciprocal component requires the antisymmetric
part of the bubble, which appears only when the pair dynamics acquires a
reactive piece, $\mathrm{Im}\,\gamma\neq0$, i.e., an imbalance between
pair-forming states above and below the Fermi level. In the expansion
\eqref{eq:expansion} the gyrotropy $g_{ijlm}k_lB_m$ is the $\vk$-linear
sibling of the Hall tensor $\lambda_{ijl}B_l$: both are governed by
$\partial T_c/\partial\mu$, and our result places the fluctuation gyrotropy
in the same class as the fluctuation Hall effect, with the
substitution of one power of $\vB$ by one power of $\vk$ weighted by the
cubic LI.

For the $C_{3v}$ invariant \eqref{eq:LI-Rashba} the same main formula
\eqref{eq:master} applies. Two differences are worth noting. First, 
the allowed linear LI cancels from the nonlocal conductivity at linear order in $\bm{B}$ (after the momentum shift it reenters
only through $O(B^2)$ corrections generated by the cubic term); the observable gyrotropy
at linear order is again controlled by the cubic terms. Second, since
$w_{C_{3v}}$ is an $\ell=1$ harmonic, the anomalous-vertex and propagator
channels are comparable in magnitude and partially cancel, and the response
tensor acquires an isotropic component in addition to the traceless one
[for $\vk\parallel\hat x$, $\vB\parallel\hat y$ we find
$\sigma^{\rm odd}_{xx}:\sigma^{\rm odd}_{yy}\simeq3:1$]. The frequency
functions remain $\mathcal{W}$ and $\mathcal{V}$ up to $O(1)$ coefficients.
The momentum-odd kinetic invariant discussed in
Appendix \ref{app:noise} contributes there at the same order (see the closing paragraph
of Appendix \ref{app:noise}). 

The two Kramers-Kronig partners translate into distinct optical
observables. For a two-dimensional film the two components are cleanly separated by
the thin-film transmission problem: for a wave incident at an angle
$\theta$, which carries the in-plane wave vector $k_\parallel=n_s(\omega/c)\sin\theta$
($n_s$ being the refractive index of the medium of incidence), and
for $|2\pi\sigma/c|\ll1$ the
transmission amplitude of a wave traversing the film is
$t\simeq1-2\pi\tilde\sigma(\vk,\omega)/c$, with
$\tilde\sigma=\sigma/\cos\theta$ for $s$ and $\tilde\sigma=\sigma\cos\theta$
for $p$ polarization in the free-standing case (a substrate modifies these
prefactors, but not the linear dependence on $\sigma^{\rm odd}$), so counter-propagating waves are
absorbed differently,
$\Delta\mathcal{A}=\mathcal{A}(+\vk)-\mathcal{A}(-\vk)\simeq
(4\pi/c)\,\mathrm{Re}\,\sigma^{\rm odd}\propto
\eta B_z(\gamma_2/\gamma_1)\,\mathcal{W}(\omega\tau_{\rm GL})$, 
nonreciprocal directional dichroism, the linear-response counterpart of the
magnetochiral dichroism of Rikken and Raupach \cite{Rikken1997}, while the
transmitted phase acquires the direction-odd shift
$\Delta\varphi=-(2\pi/c)\,\mathrm{Im}\,\sigma^{\rm odd}
\propto\mathcal{V}(\omega\tau_{\rm GL})$, gyrotropic birefringence.
Equivalently, in the longitudinal channel the two-dimensional dielectric
function is $\varepsilon(\vk,\omega)=1+2\pi ik\,\sigma_L(\vk,\omega)/\omega$.
Interferometric or polarimetric detection of $\Delta\varphi$, or
direction-resolved absorption measurements of $\Delta\mathcal{A}$, at
oblique incidence in gated
MoS$_2$ or NbSe$_2$ films near $T_c$ would constitute a direct observation
of fluctuation gyrotropy; since the effect vanishes at normal incidence
and $k_\parallel\xi\sim(\omega\xi/c)\sin\theta$ is small, near-field probes
with a definite in-plane wave vector are the more sensitive option.

The same band structure that generates the LI produces a normal-state
gyrotropy, which sets the background against which the critical enhancement
must be detected. In Appendix~\ref{app:normal} we evaluate it for the
minimal $D_{3h}$ band model of MoS$_2$ \cite{Wakatsuki2017} from the Kubo
bubble with intraband disorder. The result [Eq.~\eqref{eq:NSfinal}] has the
same traceless trigonal tensor structure \eqref{eq:tensor} and the
same microscopic combination $\lambda\Delta_{\rm SO}\Delta_Z$ as the
fluctuation contribution, but the frequency dependence is a smooth
Drude-Lorentzian in $\omega\tau$ and the magnitude
$\propto e^{2}\tau^{2}\lambda\Delta_{\rm SO}\Delta_Z$ is temperature
independent. The fluctuation-to-normal ratio thus diverges as
$1/\epsilon$ for $\omega\tau_{\rm GL}\lesssim1$: the experimental fingerprint of the effect discussed here is a
critical upturn of the nonreciprocal dichroism and birefringence on top of a
smooth background as $T\to T_c$, in direct analogy with the giant
enhancement of the MCA observed in the same regime \cite{Wakatsuki2017}.
Note also the consistency with the normal-state analysis of
Ref.~\cite{Wakatsuki2017}, which found no normal-state nonlinear
MCA in this model: the linear $\vk$-resolved gyrotropy is a distinct
(spatial-dispersion) response and does not vanish; interestingly, it is
determined by band curvature, the normal-state counterpart of the particle-hole
asymmetry factor $\gamma_2$.

Finally, the scalar AL bubble generates only the symmetric gyrotropic
tensor $g$; the antisymmetric natural-activity tensor $\chi$ vanishes
identically in this framework. This mirrors the normal-state dictionary in
which $\chi$ is tied to the orbital magnetic moment of quasiparticles
\cite{Ma2015,Zhong2016,Shinada2024}: a fluctuation contribution to natural
optical activity requires an intrinsic orbital moment of the Cooper pair,
i.e., an interband (multiband) generalization of the pair field beyond the
single scalar $\Psi$, an interesting problem for future work.

The broader message of this work is that the familiar hierarchy of
geometric linear-response coefficients (Hall, natural activity,
gyrotropy) is critically enhanced, in the fluctuation
regime of noncentrosymmetric superconductors, with the Lifshitz invariants
of the GL functional playing the role that band geometry plays in the
normal state. Several directions follow naturally. The Maki-Thompson and
density-of-states channels carry the same symmetric vertex structure and
should be evaluated, particularly in the systems where pair breaking
is weak. A microscopic (via Keldysh technique \cite{Levchenko2007}) derivation would fix the precise value of
$\gamma_2$ and of the $\eta$ coefficient of the cubic LI for the TMD band structure, and its sensitivity to disorder strength. 
On the experimental side, beyond far-field optics, the
nonreciprocal conductivity at finite $\vk$ is exactly the object probed by
near-field and quantum-sensing techniques: single-spin noise magnetometry
already measures $\mathrm{Re}\,\sigma(\vk,\omega)$ of fluctuating
superconductors at $k\sim1/z_{\rm NV}$
\cite{Liu2025}. An extension sensitive to the direction-odd
component requires a probe that breaks the in-plane isotropy of the film:
for the traceless trigonal tensor \eqref{eq:tensor} the single-point noise
of a spin above a homogeneous film averages to zero over the direction of
$\vk$, but two-point (cross-relaxation) correlations of two spins separated
by $\bm{d}$, which weight the odd sector by $\cos3\varphi_{\bm d}$, or fields
with a definite in-plane wave vector, do resolve it; in the $C_{3v}$
geometry the isotropic part of $\sigma^{\rm odd}$ produces circularly
polarized magnetic noise and hence a field-odd difference between the
$0\to\pm1$ relaxation rates of a single spin quantized along the field. In
all cases the fluctuation gyrotropy is distinguished from the Hall response
(even in $\vk$, antisymmetric in $ij$), from the normal-state gyrotropy
(temperature independent, with the Drude scale $1/\tau$), and from geometric
artifacts (even in $\vB$) by being odd in $\vB$ and in $\vk$, symmetric and
traceless with threefold angular dependence, so that it reverses sign under
a $60^\circ$ rotation of the sample, and by its critical temperature
dependence. More broadly, moir\'e and strained TMD superlattices offer
in-situ control of the trigonal warping and hence of the cubic Lifshitz
invariant, suggesting a tunable platform where the
nonreciprocal linear response, the diode effect, and the MCA can be tuned
by the same knob and compared quantitatively.

\begin{acknowledgments}

The author is grateful to Ilya Esterlis for carefully reading the manuscript and providing valuable comments, and to Andrey Varlamov for useful communication and bringing Refs.~\cite{Barash1993,Galaktionov1995} to my attenion.
This work was supported by the National Science Foundation Grant No. DMR-2452658 and H. I. Romnes Faculty Fellowship provided by the University of Wisconsin-Madison Office of the Vice Chancellor for Research and Graduate Education with funding from the Wisconsin Alumni Research Foundation. The author acknowledges the use of Claude (Anthropic) for checks with analytical derivations, their independent numerical verification, and manuscript preparation. The author conceived the project and carried research.
\end{acknowledgments}

\appendix

\section{Normal-state gyrotropy of the $D_{3h}$ band model}
\label{app:normal}

Following Ref. \cite{Wakatsuki2017}, the minimal band
model of gated MoS$_2$ combines a quadratic dispersion, trigonal warping,
Ising spin-orbit splitting, and the Zeeman energy (valley index $v=\pm1$,
spin $s=\pm1$):
\begin{equation}
\label{eq:NSbands}
E_{vs}(\vp)=\xi_{\vp}+v\lambda\,(p_x^{3}-3p_xp_y^{2})
+vs\,\Delta_{\rm SO}+s\,\Delta_Z ,
\end{equation}
with $\xi_{\vp}=p^{2}/2m-\mu$ and $\Delta_Z=\tfrac12 g\mu_B B_z$. The current vertex is diagonal,
$v_i^{vs}=\partial_{p_i}E_{vs}$, and the conductivity bubble is purely
intraband. Integrating out the pairing fluctuations of this model generates
the cubic LI \eqref{eq:LI-D3h} with $\eta B_z\propto\lambda\Delta_{\rm SO}\Delta_Z/T_c^{2}$.

Intraband disorder enters through the retarded and advanced Green's functions  
$G^{R/A}(\vp,\ep)=[\ep-E_{vs}(\vp)\pm i/2\tau]^{-1}$. Evaluating the
current-current bubble in the Matsubara representation, continuing analytically 
$i\Omega_m\to\omega+i0$, and keeping the dominant $G^RG^A$ term, the
dissipative conductivity takes the spectral (Kubo--Greenwood) form
$\propto\int d\ep\,[f(\ep)-f(\ep+\omega)]\,
\mathrm{Im}\,G^{R}(\vp_-,\ep)\,\mathrm{Im}\,G^{R}(\vp_+,\ep+\omega)$. The
radial energy integration is an overlap of two Lorentzians whose widths add,
and the bubble collapses to the collision-broadened kernel
\begin{equation}
\label{eq:NSkernel}
\sigma_{ij}(\vk,\omega)=e^{2}\!\sum_{v,s}\!\int\!\frac{d^{2}p}{(2\pi)^{2}}\,
\frac{v_i v_j\,(-\partial_E f)}
{1/\tau-i\omega+i\,\vk\cdot\vv} ,
\end{equation}
equivalent to the relaxation-time Boltzmann solution, since the denominator in the expression above is simply the resolvent of the Liouville operator. 
Expanding to linear order in $\vk$ gives,
\begin{align}
\label{eq:NSM}
&\sigma^{\rm odd}_{ij}(\vk,\omega)=-ie^{2}\,
\frac{M_{ijl}\,k_l}{(1/\tau-i\omega)^{2}},\nonumber \\
&M_{ijl}=-\!\sum_{v,s}\!\int_{\vp} v_i v_j v_l\,\partial_Ef,
\end{align}
a Fermi-surface third-velocity moment.

Extracting the part of $M$ linear in $\Delta_Z$, only two terms survive the
valley and spin summations at order $\lambda\Delta_{\rm SO}\Delta_Z$: the
anomalous-vertex term [$\partial_{\vp} w$ in $v_i$, paired with the
$(vs\Delta_{\rm SO})(s\Delta_Z)$ cross term of the occupation expansion] and
the propagator term [$w$ in the occupation argument, at third order].
Carrying out the angular averages and the radial integrals by parts, the
two channels contribute in the ratio $3:(-1)$ [the same ratio as in the AL
calculation], with the total $(\hbar=2m=1)$
\begin{equation}
\label{eq:NSMresult}
M_{xxx}=-M_{yyx}=-M_{xyy}
=\frac{12}{\pi}\,\lambda\,\Delta_{\rm SO}\,\Delta_Z,
\end{equation}
with only exponentially small thermal corrections. Restoring units
[in wavevector convention, where the warping constant $\lambda_W$ has
the dimension of energy$\times$length$^3$ and $\Lambda\equiv\lambda_W k_F^{3}/E_F$ is the relative warping,
$\Gamma=\hbar/\tau$]:
\begin{equation}
\label{eq:NSfinal}
\sigma^{\rm odd}_{xx}(\vk,\omega)
=-\frac{12}{\pi}\,\frac{e^{2}}{\hbar}\,
\Lambda\,\frac{k_x}{k_F}\,
\frac{\Delta_{\rm SO}\Delta_Z}{\Gamma^{2}}\,
\frac{i}{(1-i\omega\tau)^{2}} .
\end{equation}
The real part (dichroism) $\propto2\omega\tau/(1+\omega^{2}\tau^{2})^{2}$
is odd in $\omega$ and vanishes at dc; the imaginary part (birefringence)
$\propto(1-\omega^{2}\tau^{2})/(1+\omega^{2}\tau^{2})^{2}$ is even and
finite at dc, the same dichotomy as in the fluctuation response, but
with the impurity scale $1/\tau$ replacing the critical scale
$1/\tau_{\rm GL}$, and no temperature enhancement.

\begin{figure}[t]
\centering
\includegraphics[width=\columnwidth]{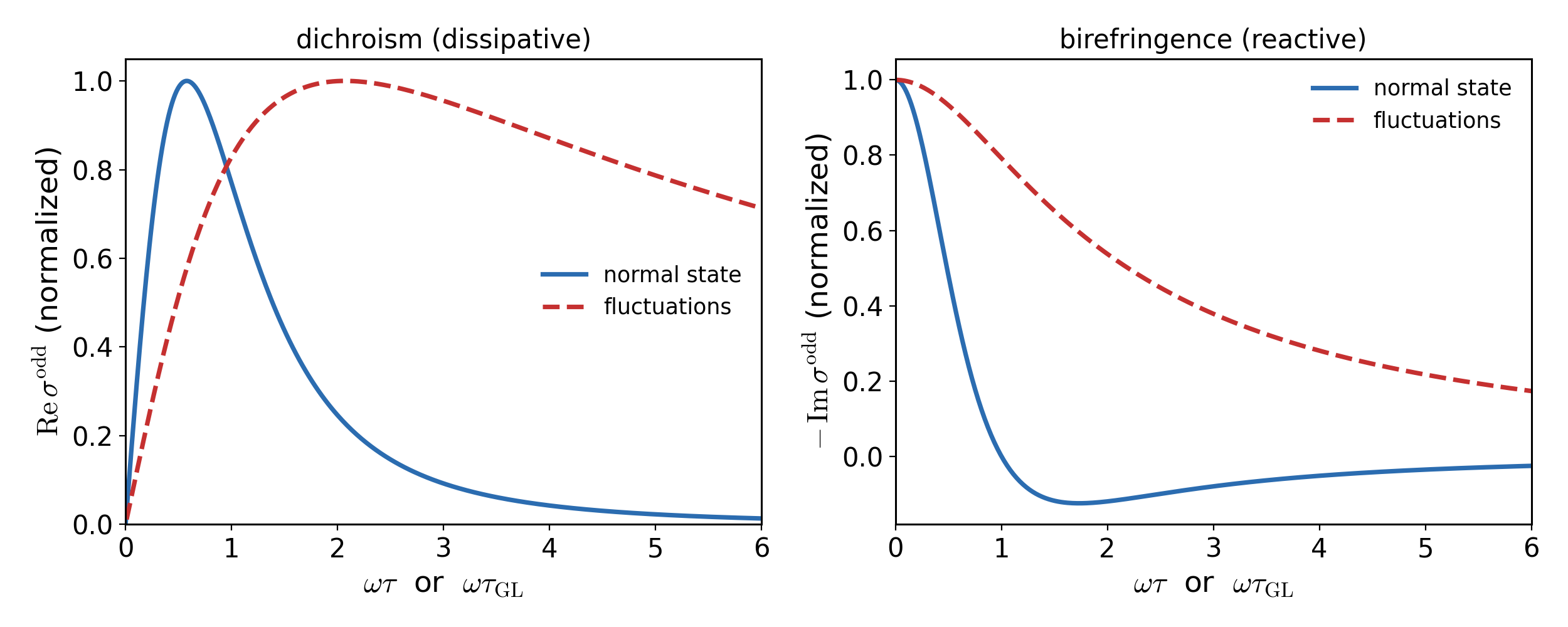}
\caption{Normalized frequency profiles of the nonreciprocal conductivity in
the normal state (solid blue, argument $\omega\tau$) and in the fluctuation
regime (dashed red, argument $\omega\tau_{\rm GL}$). Left: dichroism,
$\mathrm{Re}\,\sigma^{\rm odd}$. The normal-state curve is
$2\omega\tau/(1+\omega^{2}\tau^{2})^{2}$ normalized by its maximum
$3\sqrt3/8\simeq0.6495$ at $\omega\tau=1/\sqrt3$. The fluctuation curve is
$\mathcal{W}(\varpi)/\mathcal{W}_{\max}$ with
$\mathcal{W}_{\max}\simeq0.886$ at $\varpi\simeq2.07$. Right: birefringence,
$-\mathrm{Im}\,\sigma^{\rm odd}$. The normal-state curve is
$(1-\omega^{2}\tau^{2})/(1+\omega^{2}\tau^{2})^{2}$, which equals unity at
$\omega=0$. The fluctuation curve is $\mathcal{V}(\varpi)/\mathcal{V}(0)$
with $\mathcal{V}(0)=2$. The dimensionful prefactors removed by these
normalizations are $(12/\pi)(e^{2}/\hbar)\Lambda(k_x/k_F)
\Delta_{\rm SO}\Delta_Z/\Gamma^{2}$ for the normal state
[Eq.~\eqref{eq:NSfinal}] and
$(2e)^{2}T\gamma_2m\eta B_zk_x/8\pi a$ for the fluctuations
[Eq.~\eqref{eq:mainresult}].}
\label{fig:compare}
\end{figure}

The kernel \eqref{eq:NSkernel} is the bare bubble, and one may ask whether
impurity-ladder vertex corrections, which for the ordinary Drude
conductivity convert the quantum lifetime into the transport time, 
modify the gyrotropic response. For point-like impurities each ladder rung
carries no momentum dependence and transmits only the angular $\ell=0$
(density) harmonic of the vertex. Summing the ladder is then exactly
equivalent to solving the kinetic equation with the number-conserving 
collision integral \cite{BGK1954}, $\mathrm{St}[\delta f]=-[\delta f-(-\partial_Ef)\,\delta\mu]/\tau$,
in which collisions relax the distribution toward a local equilibrium with a
shifted chemical potential $\delta\mu$ fixed by particle
conservation. Solving the linearized kinetic equation, all response
coefficients are expressed through moments of a single scalar kernel,
\begin{align}\label{eq:Amoments}
&R(\vp)=\frac{(-\partial_{E}f)}{1/\tau-i\omega+i\vk\cdot\vv(\vp)},\quad A_{0}=\sum_{v,s}\int_{\vp} R,\, \nonumber 
\\
&A_{i}=\sum_{v,s}\int_{\vp} v_{i}R,\quad
A_{ij}=\sum_{v,s}\int_{\vp} v_{i}v_{j}R,
\end{align}
with $\int_{\vp}\equiv\int d^{2}p/(2\pi)^{2}$: a scalar (density-density
bubble), a vector (mixed current-density bubble), and a rank-two tensor
(the bare conductivity bubble), respectively. Eliminating $\delta\mu$ via
$\int\delta f=\nu_{0}\,\delta\mu$, where $\nu_{0}=\sum_{v,s}\int_{\vp}
(-\partial_{E}f)$ is the thermally smeared density of states, yields
\begin{equation}
\label{eq:BGKpaper}
\sigma_{ij}(\vk,\omega)=e^{2}\bigg[A_{ij}
+\frac{A_{i}A_{j}}{\nu_{0}\tau-A_{0}}\bigg].
\end{equation}
The second term is the ladder (diffuson) correction: the field feeds the
density channel through the vector moment $A_{j}$, and the density
channel feeds back into the current through $A_{i}$, so the correction is
the outer product of two rank-one objects divided by the scalar diffuson
denominator, $\nu_{0}\tau-A_{0}\simeq\nu_{0}\tau^{2}(-i\omega+Dk^{2})$ at
$\omega\tau\ll1$. At $\vk=0$ the vector moment vanishes by parity,
$A_{i}\equiv0$, so the Drude conductivity is unrenormalized,
$\tau_{\rm tr}=\tau$, for isotropic scattering. At finite $\vk$ the ladder is
essential in the reciprocal longitudinal channel, where it restores
charge conservation, $\sigma_{L}\to\varsigma_{xx}(-i\omega)/(-i\omega+Dk^{2})$,
a property the bare bubble violates. For the gyrotropic part,
however, angular selection rules protect the result \eqref{eq:NSfinal}: the
$\lambda\Delta_{\rm SO}\Delta_{Z}$ content of $A_{i}$ vanishes at $O(k)$
[$\langle\cos\theta\cos2\theta\rangle=
\langle\cos^{2}\theta\cos3\theta\rangle=0$], and $A_{y}\equiv0$ for
$\vk\parallel\hat{x}$ by parity. Consequently the transverse gyrotropic
component is exactly free of vertex corrections, while the
longitudinal-index components are dressed only through the diffuson pole,
with relative corrections of order $Dk^{2}/\omega$. 
Since a propagating wave has $k=n\omega/c$, so that
$Dk^{2}/\omega\sim Dn^{2}\omega/c^{2}\ll1$, all dichroism and
birefringence results of this paper are unaffected by the ladder;
in the opposite static (near-field) limit the longitudinal
gyrotropic channel is screened out by charge conservation, while the
transverse channel (the one that sources magnetic noise in qubit
relaxometry) remains exactly unrenormalized. For finite-range
(anisotropic) disorder the ladder transmits higher angular harmonics; the
$\ell=1$ and $\ell=3$ components of the velocity vertex then acquire
distinct transport times, replacing the coefficient $12/\pi$ in
Eq.~\eqref{eq:NSMresult} by a weighted combination without changing the
tensor structure.


\section{Nonreciprocal Langevin noise and the kinetic Lifshitz invariant}
\label{app:noise}

The TDGL theory of Sec.~II treats the relaxation constant
$\gamma=\gamma_1+i\gamma_2$ as momentum independent, with the white Langevin
noise \eqref{eq:FDT}. Symmetry, however, permits, and microscopic
derivations of stochastic TDGL equations generally produce \cite{Liu2026}, a momentum-odd
correction to the kinetic coefficient itself,
\begin{equation}
\label{eq:gammaqApp}
\gamma_1\;\longrightarrow\;\gamma_1(\vq)=\gamma_1\big[1+\rho(\vq)\big],
\qquad \rho(-\vq)=-\rho(\vq),
\end{equation}
a kinetic Lifshitz invariant: the relaxation rate of a fluctuating
Cooper pair depends on the direction of its momentum. 
In a microscopic calculation it can be seen that because the Keldysh
component of the pair propagator is locked to its retarded and advanced
components by the exact equilibrium identity \cite{Levchenko2007},
$L^{-1}_K=\coth(\tfrac{\omega}{2T})\,[L^{-1}_R-L^{-1}_A]$, every
nonreciprocal term of the dissipative kernel is inherited by the noise: the
Langevin correlator \eqref{eq:FDT} is replaced by
\begin{equation}
\label{eq:FDTnr}
\langle\zeta(\vq,\nu)\,\zeta^{*}(\vq',\nu')\rangle
=2\gamma_1(\vq)\,T\,(2\pi)^{3}\,\delta(\vq-\vq')\,\delta(\nu-\nu'),
\end{equation}
so that the noise power is itself direction dependent, and no
independent nonreciprocal noise constant can appear in equilibrium. 
Symmetry classifies $\rho(\vq)$ exactly as it classifies the
static invariants $w(\vq)$: Onsager reciprocity of kinetic coefficients
requires $\gamma(\vq;\vB)=\gamma(-\vq;-\vB)$, so a $\vq$-odd term must be
simultaneously $\vB$-odd. For the Rashba geometry (polar axis
$\hat{\bm z}$, in-plane field) a linear drift is allowed and is found
microscopically, $\rho=\bm{u}\cdot\vq$ with
$\bm{u}\propto\vB\times\hat{\bm z}$; for the $D_{3h}$ geometry of
the main text a linear term is forbidden by the same argument that forbids
the linear static LI, and the leading kinetic invariant is the trigonal
cubic,
\begin{equation}
\label{eq:rhoD3hApp}
\rho(\vq)=u_3\,B_z\,(q_x^{3}-3q_xq_y^{2}),
\end{equation}
the dynamical counterpart of Eq.~\eqref{eq:LI-D3h}, generated at the same
microscopic order $\lambda\Delta_{\rm SO}\Delta_Z$; as for $\eta$, we treat
$u_3$ as a phenomenological parameter, so that the estimates below are
statements of relative order rather than of absolute magnitude. In this appendix we
show that including
Eqs.~\eqref{eq:gammaqApp}--\eqref{eq:rhoD3hApp} does not modify the results
of the main text at leading order near $T_c$, and in particular leaves all
structural conclusions intact.

With the friction and noise locked by Eq.~\eqref{eq:FDTnr}, the pair
propagator becomes
\begin{equation}
\label{eq:PiNR}
\Pi(\vq,\nu)=\frac{2\gamma_1(\vq)\,T}
{\big[\ep_{\vq}+\gamma_2\nu\big]^{2}+\gamma_1(\vq)^{2}\,\nu^{2}} .
\end{equation}
Its frequency integral is independent of both $\gamma_1(\vq)$ and $\gamma_2$ (the asymmetry only shifts the center of
the Lorentzian in $\nu$, not its area)
\begin{equation}
\int\!\frac{d\nu}{2\pi}\,\Pi(\vq,\nu)=\frac{T}{\ep_{\vq}},
\end{equation}
the factors $1+\rho$ canceling between numerator and denominator: the
equal-time (thermodynamic) fluctuations remain Gibbsian, as they must, since
static thermodynamics knows nothing about kinetic coefficients. Two
consequences are immediate. First, the free energy (and with it the
static Lifshitz invariant $w(\vq)$, the shift argument for the linear LI,
and $T_c(\vB)$) is untouched. Second, at $\omega\to0$ the reciprocal
(even-in-$\vk$) conductivity of Sec.~III, which is controlled by equal-time
statistics through the kernel \eqref{eq:K0}, receives no odd-noise
correction at linear order in $\rho$: the entire even sector, including the
scaling functions $F_{T,L}$ and the noise-magnetometry application, stands
as written.

The central structural statement of the main text --- that the odd-in-$\vk$
conductivity is controlled by particle--hole asymmetry --- might appear
questionable: one could imagine the direction-dependent noise
\eqref{eq:FDTnr} directly imprinting a direction dependence on the current
correlator even at $\gamma_2=0$. This does not happen, for the following
reason. At $\gamma_2=0$ the propagator \eqref{eq:PiNR} is even in $\nu$ for
any $\gamma_1(\vq)$. Consequently, in the frequency integral of the
current loop the substitution $\nu\to-\nu$ shows that the kernel
\begin{equation}
K=\int\!\frac{d\nu}{2\pi}\,
\Pi\big(\vq_+,\nu+\tfrac{\omega}{2}\big)\,
\Pi\big(\vq_-,\nu-\tfrac{\omega}{2}\big)
\end{equation}
is a symmetric function under the simultaneous exchange of the full
label sets $(\ep_+,\gamma_{1+})\leftrightarrow(\ep_-,\gamma_{1-})$, where
$\gamma_{1\pm}\equiv\gamma_1(\vq\pm\vk/2)$. Since $\vk\to-\vk$ performs
precisely this exchange while the vertex $v_iv_j$ is $\vk$ independent, the
conductivity remains strictly even in $\vk$ for a real kinetic
coefficient of arbitrary momentum dependence, FDT-locked noise
included. The no-go argument of Sec.~IV is thus strengthened
rather than circumvented: nonreciprocal friction and noise alone cannot
generate a nonreciprocal linear response, and particle--hole
asymmetry $\gamma_2$ remains a necessary ingredient. 

Once $\gamma_2\neq0$, the kinetic invariant does open an additional channel
for the odd conductivity, in complete analogy with the static LI: expanding
the kernel to linear order in $\gamma_2$ and in the antisymmetric
combinations $\ep_+-\ep_-$ and $\gamma_{1+}-\gamma_{1-}$ produces, besides
Eq.~\eqref{eq:deltaK}, a cross term $\propto\gamma_2\,\vk\cdot
\bm{\nabla}_{\vq}\rho$. Its angular structure is fixed by symmetry: the
cubic harmonic \eqref{eq:rhoD3hApp} combined with one power of $\vk$ feeds
the same symmetric, traceless, trigonal tensor
$(\sigma_{xx}-\sigma_{yy})+2i\sigma_{xy}\propto B_z(k_x-ik_y)$ as
Eq.~\eqref{eq:mainresult},
and the parities in frequency (dissipative part odd in $\omega$ and
vanishing at dc, reactive part even and finite at dc) follow from the
reality condition $\sigma_{ij}(\vk,\omega)^{*}=\sigma_{ij}(-\vk,-\omega)$
independently of which invariant drives the response.

The two channels differ, however, in their temperature dependence. Both
invariants are cubic in the pair momentum and are therefore of order
$q^{3}\sim\xi^{-3}\propto\epsilon^{3/2}$ at the fluctuation scale; but the
static invariant competes with the pair energy,
$w/\ep_{\vq}\sim\eta B_z\xi^{-3}/a\propto\epsilon^{1/2}$, whereas the
kinetic one competes with the temperature-independent friction,
$\rho\sim u_3B_z\xi^{-3}\propto\epsilon^{3/2}$. The kinetic channel is
therefore smaller by one power of
$\epsilon$: at fixed $\varpi=\omega\tau_{\rm GL}$ the static-LI
contribution scales as $1/\epsilon$ [Eq.~\eqref{eq:mainresult}] while the
kinetic-LI contribution saturates to an $\epsilon$-independent value. 
The two channels also generically enter with independent signs, so
partial compensation is possible away from the critical region. All
leading singular results of the main text (the dichroism
$\propto\omega/\epsilon^{2}$ with the frequency function
$\mathcal{W}(\varpi)$, the birefringence $\propto1/\epsilon$ with
$\mathcal{V}(0)=2$, and the fluctuation-to-normal-state enhancement)
are thus unaffected by the kinetic Lifshitz invariant, whose contribution
constitutes a relative $O(\epsilon)$ correction with its own $O(1)$
frequency profile. The two channels become comparable only when
$\rho\sim w/\ep_{\vq}$, i.e., at $\epsilon\sim\eta/\alpha u_3$; since $\eta$
and $u_3$ share the same microscopic origin and differ by one energy scale of
order $T_c$, this is $\epsilon\sim O(1)$, outside the critical region. Pair
breaking, which shifts $\epsilon\to\epsilon+\delta$ in both channels alike,
does not change this hierarchy, so within the Gaussian regime the $1/\epsilon$
law is limited only by the fixed-frequency crossover discussed after
Eq.~\eqref{eq:staticbiref} and by the Ginzburg cutoff.

We note one qualitative difference in the Rashba geometry: there the
linear kinetic drift $\bm{u}\cdot\vq$ is allowed and, unlike the
linear static LI, is not removable by the momentum shift (a constant
shift leaves the odd part of $\gamma_1(\vq)$ invariant). Its relative
strength at the fluctuation scale, $u\,\xi^{-1}\propto\epsilon^{1/2}$,
matches that of the cubic static invariant, so in Rashba superconductors
the linear-response gyrotropy receives comparable additive contributions
from both: the linear-response counterpart of the additive drift and
cubic-gradient contributions found for the fluctuation MCA.

\bibliography{biblio-gyrotropy}

\begin{thebibliography}{60}%
\makeatletter
\providecommand \@ifxundefined [1]{%
 \@ifx{#1\undefined}
}%
\providecommand \@ifnum [1]{%
 \ifnum #1\expandafter \@firstoftwo
 \else \expandafter \@secondoftwo
 \fi
}%
\providecommand \@ifx [1]{%
 \ifx #1\expandafter \@firstoftwo
 \else \expandafter \@secondoftwo
 \fi
}%
\providecommand \natexlab [1]{#1}%
\providecommand \enquote  [1]{``#1''}%
\providecommand \bibnamefont  [1]{#1}%
\providecommand \bibfnamefont [1]{#1}%
\providecommand \citenamefont [1]{#1}%
\providecommand \href@noop [0]{\@secondoftwo}%
\providecommand \href [0]{\begingroup \@sanitize@url \@href}%
\providecommand \@href[1]{\@@startlink{#1}\@@href}%
\providecommand \@@href[1]{\endgroup#1\@@endlink}%
\providecommand \@sanitize@url [0]{\catcode `\\12\catcode `\$12\catcode
  `\&12\catcode `\#12\catcode `\^12\catcode `\_12\catcode `\%12\relax}%
\providecommand \@@startlink[1]{}%
\providecommand \@@endlink[0]{}%
\providecommand \url  [0]{\begingroup\@sanitize@url \@url }%
\providecommand \@url [1]{\endgroup\@href {#1}{\urlprefix }}%
\providecommand \urlprefix  [0]{URL }%
\providecommand \Eprint [0]{\href }%
\providecommand \doibase [0]{http://dx.doi.org/}%
\providecommand \selectlanguage [0]{\@gobble}%
\providecommand \bibinfo  [0]{\@secondoftwo}%
\providecommand \bibfield  [0]{\@secondoftwo}%
\providecommand \translation [1]{[#1]}%
\providecommand \BibitemOpen [0]{}%
\providecommand \bibitemStop [0]{}%
\providecommand \bibitemNoStop [0]{.\EOS\space}%
\providecommand \EOS [0]{\spacefactor3000\relax}%
\providecommand \BibitemShut  [1]{\csname bibitem#1\endcsname}%
\let\auto@bib@innerbib\@empty
\bibitem [{\citenamefont {Tokura}\ and\ \citenamefont
  {Nagaosa}(2018)}]{Tokura2018}%
  \BibitemOpen
  \bibfield  {author} {\bibinfo {author} {\bibfnamefont {Yoshinori}\
  \bibnamefont {Tokura}}\ and\ \bibinfo {author} {\bibfnamefont {Naoto}\
  \bibnamefont {Nagaosa}},\ }\bibfield  {title} {\enquote {\bibinfo {title}
  {Nonreciprocal responses from non-centrosymmetric quantum materials},}\
  }\href {\doibase 10.1038/s41467-018-05759-4} {\bibfield  {journal} {\bibinfo
  {journal} {Nat. Commun.}\ }\textbf {\bibinfo {volume} {9}},\ \bibinfo {pages}
  {3740} (\bibinfo {year} {2018})}\BibitemShut {NoStop}%
\bibitem [{\citenamefont {Ideue}\ and\ \citenamefont
  {Iwasa}(2021)}]{Ideue2021}%
  \BibitemOpen
  \bibfield  {author} {\bibinfo {author} {\bibfnamefont {Toshiya}\ \bibnamefont
  {Ideue}}\ and\ \bibinfo {author} {\bibfnamefont {Yoshihiro}\ \bibnamefont
  {Iwasa}},\ }\bibfield  {title} {\enquote {\bibinfo {title} {Symmetry breaking
  and nonlinear electric transport in van der {Waals} nanostructures},}\ }\href
  {\doibase 10.1146/annurev-conmatphys-060220-100347} {\bibfield  {journal}
  {\bibinfo  {journal} {Annu. Rev. Condens. Matter Phys.}\ }\textbf {\bibinfo
  {volume} {12}},\ \bibinfo {pages} {201--223} (\bibinfo {year}
  {2021})}\BibitemShut {NoStop}%
\bibitem [{\citenamefont {Nagaosa}\ and\ \citenamefont
  {Yanase}(2024)}]{Nagaosa2024}%
  \BibitemOpen
  \bibfield  {author} {\bibinfo {author} {\bibfnamefont {Naoto}\ \bibnamefont
  {Nagaosa}}\ and\ \bibinfo {author} {\bibfnamefont {Youichi}\ \bibnamefont
  {Yanase}},\ }\bibfield  {title} {\enquote {\bibinfo {title} {Nonreciprocal
  transport and optical phenomena in quantum materials},}\ }\href {\doibase
  10.1146/annurev-conmatphys-032822-033734} {\bibfield  {journal} {\bibinfo
  {journal} {Annu. Rev. Condens. Matter Phys.}\ }\textbf {\bibinfo {volume}
  {15}},\ \bibinfo {pages} {63--83} (\bibinfo {year} {2024})}\BibitemShut
  {NoStop}%
\bibitem [{\citenamefont {Ando}\ \emph {et~al.}(2020)\citenamefont {Ando},
  \citenamefont {Miyasaka}, \citenamefont {Li}, \citenamefont {Ishizuka},
  \citenamefont {Arakawa}, \citenamefont {Shiota}, \citenamefont {Moriyama},
  \citenamefont {Yanase},\ and\ \citenamefont {Ono}}]{Ando2020}%
  \BibitemOpen
  \bibfield  {author} {\bibinfo {author} {\bibfnamefont {Fuyuki}\ \bibnamefont
  {Ando}}, \bibinfo {author} {\bibfnamefont {Yuta}\ \bibnamefont {Miyasaka}},
  \bibinfo {author} {\bibfnamefont {Tian}\ \bibnamefont {Li}}, \bibinfo
  {author} {\bibfnamefont {Jun}\ \bibnamefont {Ishizuka}}, \bibinfo {author}
  {\bibfnamefont {Tomonori}\ \bibnamefont {Arakawa}}, \bibinfo {author}
  {\bibfnamefont {Yoichi}\ \bibnamefont {Shiota}}, \bibinfo {author}
  {\bibfnamefont {Takahiro}\ \bibnamefont {Moriyama}}, \bibinfo {author}
  {\bibfnamefont {Youichi}\ \bibnamefont {Yanase}}, \ and\ \bibinfo {author}
  {\bibfnamefont {Teruo}\ \bibnamefont {Ono}},\ }\bibfield  {title} {\enquote
  {\bibinfo {title} {Observation of superconducting diode effect},}\ }\href
  {\doibase 10.1038/s41586-020-2590-4} {\bibfield  {journal} {\bibinfo
  {journal} {Nature}\ }\textbf {\bibinfo {volume} {584}},\ \bibinfo {pages}
  {373} (\bibinfo {year} {2020})}\BibitemShut {NoStop}%
\bibitem [{\citenamefont {Daido}\ \emph {et~al.}(2022)\citenamefont {Daido},
  \citenamefont {Ikeda},\ and\ \citenamefont {Yanase}}]{Daido2022}%
  \BibitemOpen
  \bibfield  {author} {\bibinfo {author} {\bibfnamefont {Akito}\ \bibnamefont
  {Daido}}, \bibinfo {author} {\bibfnamefont {Yuhei}\ \bibnamefont {Ikeda}}, \
  and\ \bibinfo {author} {\bibfnamefont {Youichi}\ \bibnamefont {Yanase}},\
  }\bibfield  {title} {\enquote {\bibinfo {title} {Intrinsic superconducting
  diode effect},}\ }\href {\doibase 10.1103/PhysRevLett.128.037001} {\bibfield
  {journal} {\bibinfo  {journal} {Phys. Rev. Lett.}\ }\textbf {\bibinfo
  {volume} {128}},\ \bibinfo {pages} {037001} (\bibinfo {year}
  {2022})}\BibitemShut {NoStop}%
\bibitem [{\citenamefont {Yuan}\ and\ \citenamefont {Fu}(2022)}]{YuanFu2022}%
  \BibitemOpen
  \bibfield  {author} {\bibinfo {author} {\bibfnamefont {Noah F.~Q.}\
  \bibnamefont {Yuan}}\ and\ \bibinfo {author} {\bibfnamefont {Liang}\
  \bibnamefont {Fu}},\ }\bibfield  {title} {\enquote {\bibinfo {title}
  {Supercurrent diode effect and finite-momentum superconductors},}\ }\href
  {\doibase 10.1073/pnas.2119548119} {\bibfield  {journal} {\bibinfo  {journal}
  {Proc. Natl. Acad. Sci. USA}\ }\textbf {\bibinfo {volume} {119}},\ \bibinfo
  {pages} {e2119548119} (\bibinfo {year} {2022})}\BibitemShut {NoStop}%
\bibitem [{\citenamefont {Ili{\'c}}\ and\ \citenamefont
  {Bergeret}(2022)}]{Ilic2022}%
  \BibitemOpen
  \bibfield  {author} {\bibinfo {author} {\bibfnamefont {S.}~\bibnamefont
  {Ili{\'c}}}\ and\ \bibinfo {author} {\bibfnamefont {F.~S.}\ \bibnamefont
  {Bergeret}},\ }\bibfield  {title} {\enquote {\bibinfo {title} {Theory of the
  supercurrent diode effect in {Rashba} superconductors with arbitrary
  disorder},}\ }\href {\doibase 10.1103/PhysRevLett.128.177001} {\bibfield
  {journal} {\bibinfo  {journal} {Phys. Rev. Lett.}\ }\textbf {\bibinfo
  {volume} {128}},\ \bibinfo {pages} {177001} (\bibinfo {year}
  {2022})}\BibitemShut {NoStop}%
\bibitem [{\citenamefont {Nadeem}\ \emph {et~al.}(2023)\citenamefont {Nadeem},
  \citenamefont {Fuhrer},\ and\ \citenamefont {Wang}}]{Nadeem2023}%
  \BibitemOpen
  \bibfield  {author} {\bibinfo {author} {\bibfnamefont {Muhammad}\
  \bibnamefont {Nadeem}}, \bibinfo {author} {\bibfnamefont {Michael~S.}\
  \bibnamefont {Fuhrer}}, \ and\ \bibinfo {author} {\bibfnamefont {Xiaolin}\
  \bibnamefont {Wang}},\ }\bibfield  {title} {\enquote {\bibinfo {title} {The
  superconducting diode effect},}\ }\href {\doibase 10.1038/s42254-023-00632-w}
  {\bibfield  {journal} {\bibinfo  {journal} {Nat. Rev. Phys.}\ }\textbf
  {\bibinfo {volume} {5}},\ \bibinfo {pages} {558--577} (\bibinfo {year}
  {2023})}\BibitemShut {NoStop}%
\bibitem [{\citenamefont {Shaffer}\ and\ \citenamefont
  {Levchenko}(2025)}]{Shaffer2025}%
  \BibitemOpen
  \bibfield  {author} {\bibinfo {author} {\bibfnamefont {Daniel}\ \bibnamefont
  {Shaffer}}\ and\ \bibinfo {author} {\bibfnamefont {Alex}\ \bibnamefont
  {Levchenko}},\ }\href {https://arxiv.org/abs/2510.25864} {\enquote {\bibinfo
  {title} {Theories of superconducting diode effects},}\ } (\bibinfo {year}
  {2025}),\ \Eprint {http://arxiv.org/abs/2510.25864} {arXiv:2510.25864
  [cond-mat.supr-con]} \BibitemShut {NoStop}%
\bibitem [{\citenamefont {Rikken}\ \emph {et~al.}(2001)\citenamefont {Rikken},
  \citenamefont {F{\"o}lling},\ and\ \citenamefont {Wyder}}]{Rikken2001}%
  \BibitemOpen
  \bibfield  {author} {\bibinfo {author} {\bibfnamefont {G.~L. J.~A.}\
  \bibnamefont {Rikken}}, \bibinfo {author} {\bibfnamefont {J.}~\bibnamefont
  {F{\"o}lling}}, \ and\ \bibinfo {author} {\bibfnamefont {P.}~\bibnamefont
  {Wyder}},\ }\bibfield  {title} {\enquote {\bibinfo {title} {Electrical
  magnetochiral anisotropy},}\ }\href {\doibase 10.1103/PhysRevLett.87.236602}
  {\bibfield  {journal} {\bibinfo  {journal} {Phys. Rev. Lett.}\ }\textbf
  {\bibinfo {volume} {87}},\ \bibinfo {pages} {236602} (\bibinfo {year}
  {2001})}\BibitemShut {NoStop}%
\bibitem [{\citenamefont {Rikken}\ and\ \citenamefont
  {Raupach}(1997)}]{Rikken1997}%
  \BibitemOpen
  \bibfield  {author} {\bibinfo {author} {\bibfnamefont {G.~L. J.~A.}\
  \bibnamefont {Rikken}}\ and\ \bibinfo {author} {\bibfnamefont
  {E.}~\bibnamefont {Raupach}},\ }\bibfield  {title} {\enquote {\bibinfo
  {title} {Observation of magneto-chiral dichroism},}\ }\href {\doibase
  10.1038/37323} {\bibfield  {journal} {\bibinfo  {journal} {Nature}\ }\textbf
  {\bibinfo {volume} {390}},\ \bibinfo {pages} {493} (\bibinfo {year}
  {1997})}\BibitemShut {NoStop}%
\bibitem [{\citenamefont {Wakatsuki}\ \emph {et~al.}(2017)\citenamefont
  {Wakatsuki}, \citenamefont {Saito}, \citenamefont {Hoshino}, \citenamefont
  {Itahashi}, \citenamefont {Ideue}, \citenamefont {Ezawa}, \citenamefont
  {Iwasa},\ and\ \citenamefont {Nagaosa}}]{Wakatsuki2017}%
  \BibitemOpen
  \bibfield  {author} {\bibinfo {author} {\bibfnamefont {Ryohei}\ \bibnamefont
  {Wakatsuki}}, \bibinfo {author} {\bibfnamefont {Yu}~\bibnamefont {Saito}},
  \bibinfo {author} {\bibfnamefont {Shintaro}\ \bibnamefont {Hoshino}},
  \bibinfo {author} {\bibfnamefont {Yuki~M.}\ \bibnamefont {Itahashi}},
  \bibinfo {author} {\bibfnamefont {Toshiya}\ \bibnamefont {Ideue}}, \bibinfo
  {author} {\bibfnamefont {Motohiko}\ \bibnamefont {Ezawa}}, \bibinfo {author}
  {\bibfnamefont {Yoshihiro}\ \bibnamefont {Iwasa}}, \ and\ \bibinfo {author}
  {\bibfnamefont {Naoto}\ \bibnamefont {Nagaosa}},\ }\bibfield  {title}
  {\enquote {\bibinfo {title} {Nonreciprocal charge transport in
  noncentrosymmetric superconductors},}\ }\href {\doibase
  10.1126/sciadv.1602390} {\bibfield  {journal} {\bibinfo  {journal} {Sci.
  Adv.}\ }\textbf {\bibinfo {volume} {3}},\ \bibinfo {pages} {e1602390}
  (\bibinfo {year} {2017})}\BibitemShut {NoStop}%
\bibitem [{\citenamefont {Itahashi}\ \emph {et~al.}(2020)\citenamefont
  {Itahashi}, \citenamefont {Ideue}, \citenamefont {Saito}, \citenamefont
  {Shiogai}, \citenamefont {Nojima},\ and\ \citenamefont
  {Iwasa}}]{Itahashi2020}%
  \BibitemOpen
  \bibfield  {author} {\bibinfo {author} {\bibfnamefont {Yuki~M.}\ \bibnamefont
  {Itahashi}}, \bibinfo {author} {\bibfnamefont {Toshiya}\ \bibnamefont
  {Ideue}}, \bibinfo {author} {\bibfnamefont {Yu}~\bibnamefont {Saito}},
  \bibinfo {author} {\bibfnamefont {Junichi}\ \bibnamefont {Shiogai}}, \bibinfo
  {author} {\bibfnamefont {Tsutomu}\ \bibnamefont {Nojima}}, \ and\ \bibinfo
  {author} {\bibfnamefont {Yoshihiro}\ \bibnamefont {Iwasa}},\ }\bibfield
  {title} {\enquote {\bibinfo {title} {Nonreciprocal transport in gate-induced
  polar superconductor {SrTiO$_3$}},}\ }\href {\doibase 10.1126/sciadv.aay9120}
  {\bibfield  {journal} {\bibinfo  {journal} {Sci. Adv.}\ }\textbf {\bibinfo
  {volume} {6}},\ \bibinfo {pages} {eaay9120} (\bibinfo {year}
  {2020})}\BibitemShut {NoStop}%
\bibitem [{\citenamefont {Kovalev}\ \emph {et~al.}(2021)\citenamefont
  {Kovalev}, \citenamefont {Sonowal},\ and\ \citenamefont
  {Savenko}}]{Kovalev2021}%
  \BibitemOpen
  \bibfield  {author} {\bibinfo {author} {\bibfnamefont {V.~M.}\ \bibnamefont
  {Kovalev}}, \bibinfo {author} {\bibfnamefont {K.}~\bibnamefont {Sonowal}}, \
  and\ \bibinfo {author} {\bibfnamefont {I.~G.}\ \bibnamefont {Savenko}},\
  }\bibfield  {title} {\enquote {\bibinfo {title} {Coherent photogalvanic
  effect in fluctuating superconductors},}\ }\href {\doibase
  10.1103/PhysRevB.103.024513} {\bibfield  {journal} {\bibinfo  {journal}
  {Phys. Rev. B}\ }\textbf {\bibinfo {volume} {103}},\ \bibinfo {pages}
  {024513} (\bibinfo {year} {2021})}\BibitemShut {NoStop}%
\bibitem [{\citenamefont {Parafilo}\ \emph {et~al.}(2022)\citenamefont
  {Parafilo}, \citenamefont {Boev}, \citenamefont {Kovalev},\ and\
  \citenamefont {Savenko}}]{Parafilo2022}%
  \BibitemOpen
  \bibfield  {author} {\bibinfo {author} {\bibfnamefont {A.~V.}\ \bibnamefont
  {Parafilo}}, \bibinfo {author} {\bibfnamefont {M.~V.}\ \bibnamefont {Boev}},
  \bibinfo {author} {\bibfnamefont {V.~M.}\ \bibnamefont {Kovalev}}, \ and\
  \bibinfo {author} {\bibfnamefont {I.~G.}\ \bibnamefont {Savenko}},\
  }\bibfield  {title} {\enquote {\bibinfo {title} {Photogalvanic transport in
  fluctuating {Ising} superconductors},}\ }\href {\doibase
  10.1103/PhysRevB.106.144502} {\bibfield  {journal} {\bibinfo  {journal}
  {Phys. Rev. B}\ }\textbf {\bibinfo {volume} {106}},\ \bibinfo {pages}
  {144502} (\bibinfo {year} {2022})}\BibitemShut {NoStop}%
\bibitem [{\citenamefont {Mironov}\ \emph {et~al.}(2024)\citenamefont
  {Mironov}, \citenamefont {Mel'nikov},\ and\ \citenamefont
  {Buzdin}}]{Buzdin2024}%
  \BibitemOpen
  \bibfield  {author} {\bibinfo {author} {\bibfnamefont {S.~V.}\ \bibnamefont
  {Mironov}}, \bibinfo {author} {\bibfnamefont {A.~S.}\ \bibnamefont
  {Mel'nikov}}, \ and\ \bibinfo {author} {\bibfnamefont {A.~I.}\ \bibnamefont
  {Buzdin}},\ }\bibfield  {title} {\enquote {\bibinfo {title} {Photogalvanic
  phenomena in superconductors supporting intrinsic diode effect},}\ }\href
  {\doibase 10.1103/PhysRevB.109.L220503} {\bibfield  {journal} {\bibinfo
  {journal} {Phys. Rev. B}\ }\textbf {\bibinfo {volume} {109}},\ \bibinfo
  {pages} {L220503} (\bibinfo {year} {2024})}\BibitemShut {NoStop}%
\bibitem [{\citenamefont {de~Miranda}\ and\ \citenamefont
  {Levchenko}(2026)}]{JTM2026c}%
  \BibitemOpen
  \bibfield  {author} {\bibinfo {author} {\bibfnamefont {Joaquim~Telles}\
  \bibnamefont {de~Miranda}}\ and\ \bibinfo {author} {\bibfnamefont {Alex}\
  \bibnamefont {Levchenko}},\ }\href {https://arxiv.org/abs/2608.20166}
  {\enquote {\bibinfo {title} {Photogalvanic transport of nonreciprocal
  {Cooper}-pair fluctuations},}\ } (\bibinfo {year} {2026}),\ \Eprint
  {http://arxiv.org/abs/2608.20166} {arXiv:2608.20166 [cond-mat.supr-con]}
  \BibitemShut {NoStop}%
\bibitem [{\citenamefont {Onsager}(1931)}]{Onsager1931}%
  \BibitemOpen
  \bibfield  {author} {\bibinfo {author} {\bibfnamefont {Lars}\ \bibnamefont
  {Onsager}},\ }\bibfield  {title} {\enquote {\bibinfo {title} {Reciprocal
  relations in irreversible processes. {I.}}}\ }\href {\doibase
  10.1103/PhysRev.37.405} {\bibfield  {journal} {\bibinfo  {journal} {Phys.
  Rev.}\ }\textbf {\bibinfo {volume} {37}},\ \bibinfo {pages} {405} (\bibinfo
  {year} {1931})}\BibitemShut {NoStop}%
\bibitem [{\citenamefont {Landau}\ \emph {et~al.}(1984)\citenamefont {Landau},
  \citenamefont {Lifshitz},\ and\ \citenamefont
  {Pitaevskii}}]{LandauLifshitzECM}%
  \BibitemOpen
  \bibfield  {author} {\bibinfo {author} {\bibfnamefont {L.~D.}\ \bibnamefont
  {Landau}}, \bibinfo {author} {\bibfnamefont {E.~M.}\ \bibnamefont
  {Lifshitz}}, \ and\ \bibinfo {author} {\bibfnamefont {L.~P.}\ \bibnamefont
  {Pitaevskii}},\ }\href@noop {} {\emph {\bibinfo {title} {Electrodynamics of
  Continuous Media}}},\ \bibinfo {edition} {2nd}\ ed.\ (\bibinfo  {publisher}
  {Pergamon Press, Oxford},\ \bibinfo {year} {1984})\BibitemShut {NoStop}%
\bibitem [{\citenamefont {Hornreich}\ and\ \citenamefont
  {Shtrikman}(1968)}]{Hornreich1968}%
  \BibitemOpen
  \bibfield  {author} {\bibinfo {author} {\bibfnamefont {R.~M.}\ \bibnamefont
  {Hornreich}}\ and\ \bibinfo {author} {\bibfnamefont {S.}~\bibnamefont
  {Shtrikman}},\ }\bibfield  {title} {\enquote {\bibinfo {title} {Theory of
  gyrotropic birefringence},}\ }\href {\doibase 10.1103/PhysRev.171.1065}
  {\bibfield  {journal} {\bibinfo  {journal} {Phys. Rev.}\ }\textbf {\bibinfo
  {volume} {171}},\ \bibinfo {pages} {1065} (\bibinfo {year}
  {1968})}\BibitemShut {NoStop}%
\bibitem [{\citenamefont {Ma}\ and\ \citenamefont {Pesin}(2015)}]{Ma2015}%
  \BibitemOpen
  \bibfield  {author} {\bibinfo {author} {\bibfnamefont {Jing}\ \bibnamefont
  {Ma}}\ and\ \bibinfo {author} {\bibfnamefont {D.~A.}\ \bibnamefont {Pesin}},\
  }\bibfield  {title} {\enquote {\bibinfo {title} {Chiral magnetic effect and
  natural optical activity in metals with or without {Weyl} points},}\ }\href
  {\doibase 10.1103/PhysRevB.92.235205} {\bibfield  {journal} {\bibinfo
  {journal} {Phys. Rev. B}\ }\textbf {\bibinfo {volume} {92}},\ \bibinfo
  {pages} {235205} (\bibinfo {year} {2015})}\BibitemShut {NoStop}%
\bibitem [{\citenamefont {Zhong}\ \emph {et~al.}(2016)\citenamefont {Zhong},
  \citenamefont {Moore},\ and\ \citenamefont {Souza}}]{Zhong2016}%
  \BibitemOpen
  \bibfield  {author} {\bibinfo {author} {\bibfnamefont {Shudan}\ \bibnamefont
  {Zhong}}, \bibinfo {author} {\bibfnamefont {Joel~E.}\ \bibnamefont {Moore}},
  \ and\ \bibinfo {author} {\bibfnamefont {Ivo}\ \bibnamefont {Souza}},\
  }\bibfield  {title} {\enquote {\bibinfo {title} {Gyrotropic magnetic effect
  and the magnetic moment on the {Fermi} surface},}\ }\href {\doibase
  10.1103/PhysRevLett.116.077201} {\bibfield  {journal} {\bibinfo  {journal}
  {Phys. Rev. Lett.}\ }\textbf {\bibinfo {volume} {116}},\ \bibinfo {pages}
  {077201} (\bibinfo {year} {2016})}\BibitemShut {NoStop}%
\bibitem [{\citenamefont {Shinada}\ and\ \citenamefont
  {Peters}(2024)}]{Shinada2024}%
  \BibitemOpen
  \bibfield  {author} {\bibinfo {author} {\bibfnamefont {Koki}\ \bibnamefont
  {Shinada}}\ and\ \bibinfo {author} {\bibfnamefont {Robert}\ \bibnamefont
  {Peters}},\ }\bibfield  {title} {\enquote {\bibinfo {title} {Orbital optical
  activity in noncentrosymmetric metals and superconductors},}\ }\href
  {\doibase 10.1103/PhysRevB.110.085162} {\bibfield  {journal} {\bibinfo
  {journal} {Phys. Rev. B}\ }\textbf {\bibinfo {volume} {110}},\ \bibinfo
  {pages} {085162} (\bibinfo {year} {2024})}\BibitemShut {NoStop}%
\bibitem [{\citenamefont {Nagaosa}\ \emph {et~al.}(2010)\citenamefont
  {Nagaosa}, \citenamefont {Sinova}, \citenamefont {Onoda}, \citenamefont
  {MacDonald},\ and\ \citenamefont {Ong}}]{Nagaosa2010AHE}%
  \BibitemOpen
  \bibfield  {author} {\bibinfo {author} {\bibfnamefont {Naoto}\ \bibnamefont
  {Nagaosa}}, \bibinfo {author} {\bibfnamefont {Jairo}\ \bibnamefont {Sinova}},
  \bibinfo {author} {\bibfnamefont {Shigeki}\ \bibnamefont {Onoda}}, \bibinfo
  {author} {\bibfnamefont {A.~H.}\ \bibnamefont {MacDonald}}, \ and\ \bibinfo
  {author} {\bibfnamefont {N.~P.}\ \bibnamefont {Ong}},\ }\bibfield  {title}
  {\enquote {\bibinfo {title} {Anomalous {Hall} effect},}\ }\href {\doibase
  10.1103/RevModPhys.82.1539} {\bibfield  {journal} {\bibinfo  {journal} {Rev.
  Mod. Phys.}\ }\textbf {\bibinfo {volume} {82}},\ \bibinfo {pages} {1539}
  (\bibinfo {year} {2010})}\BibitemShut {NoStop}%
\bibitem [{\citenamefont {Xiao}\ \emph {et~al.}(2010)\citenamefont {Xiao},
  \citenamefont {Chang},\ and\ \citenamefont {Niu}}]{Niu2010}%
  \BibitemOpen
  \bibfield  {author} {\bibinfo {author} {\bibfnamefont {Di}~\bibnamefont
  {Xiao}}, \bibinfo {author} {\bibfnamefont {Ming-Che}\ \bibnamefont {Chang}},
  \ and\ \bibinfo {author} {\bibfnamefont {Qian}\ \bibnamefont {Niu}},\
  }\bibfield  {title} {\enquote {\bibinfo {title} {Berry phase effects on
  electronic properties},}\ }\href {\doibase 10.1103/RevModPhys.82.1959}
  {\bibfield  {journal} {\bibinfo  {journal} {Rev. Mod. Phys.}\ }\textbf
  {\bibinfo {volume} {82}},\ \bibinfo {pages} {1959--2007} (\bibinfo {year}
  {2010})}\BibitemShut {NoStop}%
\bibitem [{\citenamefont {Hoshino}\ \emph {et~al.}(2018)\citenamefont
  {Hoshino}, \citenamefont {Wakatsuki}, \citenamefont {Hamamoto},\ and\
  \citenamefont {Nagaosa}}]{Hoshino2018}%
  \BibitemOpen
  \bibfield  {author} {\bibinfo {author} {\bibfnamefont {Shintaro}\
  \bibnamefont {Hoshino}}, \bibinfo {author} {\bibfnamefont {Ryohei}\
  \bibnamefont {Wakatsuki}}, \bibinfo {author} {\bibfnamefont {Keita}\
  \bibnamefont {Hamamoto}}, \ and\ \bibinfo {author} {\bibfnamefont {Naoto}\
  \bibnamefont {Nagaosa}},\ }\bibfield  {title} {\enquote {\bibinfo {title}
  {Nonreciprocal charge transport in two-dimensional noncentrosymmetric
  superconductors},}\ }\href {\doibase 10.1103/PhysRevB.98.054510} {\bibfield
  {journal} {\bibinfo  {journal} {Phys. Rev. B}\ }\textbf {\bibinfo {volume}
  {98}},\ \bibinfo {pages} {054510} (\bibinfo {year} {2018})}\BibitemShut
  {NoStop}%
\bibitem [{\citenamefont {Wakatsuki}\ and\ \citenamefont
  {Nagaosa}(2018)}]{WakatsukiNagaosa2018}%
  \BibitemOpen
  \bibfield  {author} {\bibinfo {author} {\bibfnamefont {Ryohei}\ \bibnamefont
  {Wakatsuki}}\ and\ \bibinfo {author} {\bibfnamefont {Naoto}\ \bibnamefont
  {Nagaosa}},\ }\bibfield  {title} {\enquote {\bibinfo {title} {Nonreciprocal
  current in noncentrosymmetric {Rashba} superconductors},}\ }\href {\doibase
  10.1103/PhysRevLett.121.026601} {\bibfield  {journal} {\bibinfo  {journal}
  {Phys. Rev. Lett.}\ }\textbf {\bibinfo {volume} {121}},\ \bibinfo {pages}
  {026601} (\bibinfo {year} {2018})}\BibitemShut {NoStop}%
\bibitem [{\citenamefont {de~Miranda}\ \emph
  {et~al.}(2026{\natexlab{a}})\citenamefont {de~Miranda}, \citenamefont
  {Khodas},\ and\ \citenamefont {Levchenko}}]{JTM2026a}%
  \BibitemOpen
  \bibfield  {author} {\bibinfo {author} {\bibfnamefont {Joaquim~Telles}\
  \bibnamefont {de~Miranda}}, \bibinfo {author} {\bibfnamefont {Maxim}\
  \bibnamefont {Khodas}}, \ and\ \bibinfo {author} {\bibfnamefont {Alex}\
  \bibnamefont {Levchenko}},\ }\bibfield  {title} {\enquote {\bibinfo {title}
  {Magnetochiral anisotropy in strained superconducting transition metal
  dichalcogenides},}\ }\href {\doibase 10.1103/wvvl-wv6z} {\bibfield  {journal}
  {\bibinfo  {journal} {Phys. Rev. B}\ }\textbf {\bibinfo {volume} {114}},\
  \bibinfo {pages} {154503} (\bibinfo {year} {2026}{\natexlab{a}})}\BibitemShut
  {NoStop}%
\bibitem [{\citenamefont {de~Miranda}\ \emph
  {et~al.}(2026{\natexlab{b}})\citenamefont {de~Miranda}, \citenamefont
  {Khodas},\ and\ \citenamefont {Levchenko}}]{JTM2026b}%
  \BibitemOpen
  \bibfield  {author} {\bibinfo {author} {\bibfnamefont {Joaquim~Telles}\
  \bibnamefont {de~Miranda}}, \bibinfo {author} {\bibfnamefont {Maxim}\
  \bibnamefont {Khodas}}, \ and\ \bibinfo {author} {\bibfnamefont {Alex}\
  \bibnamefont {Levchenko}},\ }\bibfield  {title} {\enquote {\bibinfo {title}
  {Electrical magnetochiral anisotropy in {Rashba} superconductors},}\ }\href
  {\doibase 10.1063/10.0046092} {\bibfield  {journal} {\bibinfo  {journal} {Low
  Temperature Physics}\ }\textbf {\bibinfo {volume} {52}},\ \bibinfo {pages}
  {1107--1112} (\bibinfo {year} {2026}{\natexlab{b}})}\BibitemShut {NoStop}%
\bibitem [{\citenamefont {Daido}\ and\ \citenamefont
  {Yanase}(2024)}]{Daido2024}%
  \BibitemOpen
  \bibfield  {author} {\bibinfo {author} {\bibfnamefont {Akito}\ \bibnamefont
  {Daido}}\ and\ \bibinfo {author} {\bibfnamefont {Youichi}\ \bibnamefont
  {Yanase}},\ }\bibfield  {title} {\enquote {\bibinfo {title} {Rectification
  and nonlinear {Hall} effect by fluctuating finite-momentum {Cooper} pairs},}\
  }\href {\doibase 10.1103/PhysRevResearch.6.L022009} {\bibfield  {journal}
  {\bibinfo  {journal} {Phys. Rev. Res.}\ }\textbf {\bibinfo {volume} {6}},\
  \bibinfo {pages} {L022009} (\bibinfo {year} {2024})}\BibitemShut {NoStop}%
\bibitem [{\citenamefont {Boeva}\ and\ \citenamefont
  {Kovalev}(2024)}]{Boeva2024}%
  \BibitemOpen
  \bibfield  {author} {\bibinfo {author} {\bibfnamefont {M.~V.}\ \bibnamefont
  {Boeva}}\ and\ \bibinfo {author} {\bibfnamefont {V.~M.}\ \bibnamefont
  {Kovalev}},\ }\bibfield  {title} {\enquote {\bibinfo {title} {Photovoltaic
  {Hall} effect in two-dimensional fluctuating superconductors},}\ }\href
  {\doibase 10.1134/S002136402460294X} {\bibfield  {journal} {\bibinfo
  {journal} {JETP Letters}\ }\textbf {\bibinfo {volume} {120}},\ \bibinfo
  {pages} {494--498} (\bibinfo {year} {2024})}\BibitemShut {NoStop}%
\bibitem [{\citenamefont {Dong}\ \emph {et~al.}(2025)\citenamefont {Dong},
  \citenamefont {Yang},\ and\ \citenamefont {Zhang}}]{Dong2025}%
  \BibitemOpen
  \bibfield  {author} {\bibinfo {author} {\bibfnamefont {Zi-Hao}\ \bibnamefont
  {Dong}}, \bibinfo {author} {\bibfnamefont {Hui}\ \bibnamefont {Yang}}, \ and\
  \bibinfo {author} {\bibfnamefont {Yi}~\bibnamefont {Zhang}},\ }\bibfield
  {title} {\enquote {\bibinfo {title} {Enhanced nonlinear {Hall} effect by
  {Cooper} pairs near the superconducting phase transition},}\ }\href {\doibase
  10.1103/PhysRevB.111.155120} {\bibfield  {journal} {\bibinfo  {journal}
  {Phys. Rev. B}\ }\textbf {\bibinfo {volume} {111}},\ \bibinfo {pages}
  {155120} (\bibinfo {year} {2025})}\BibitemShut {NoStop}%
\bibitem [{\citenamefont {Mineev}\ and\ \citenamefont
  {Samokhin}(1994)}]{MineevSamokhin1994}%
  \BibitemOpen
  \bibfield  {author} {\bibinfo {author} {\bibfnamefont {V.~P.}\ \bibnamefont
  {Mineev}}\ and\ \bibinfo {author} {\bibfnamefont {K.~V.}\ \bibnamefont
  {Samokhin}},\ }\bibfield  {title} {\enquote {\bibinfo {title} {Helical phases
  in superconductors},}\ }\href@noop {} {\bibfield  {journal} {\bibinfo
  {journal} {Zh. Eksp. Teor. Fiz.}\ }\textbf {\bibinfo {volume} {105}},\
  \bibinfo {pages} {747} (\bibinfo {year} {1994})},\ \bibinfo {note} {[Sov.
  Phys. JETP {\bf 78}, 401 (1994)]}\BibitemShut {NoStop}%
\bibitem [{\citenamefont {Kaur}\ \emph {et~al.}(2005)\citenamefont {Kaur},
  \citenamefont {Agterberg},\ and\ \citenamefont {Sigrist}}]{Kaur2005}%
  \BibitemOpen
  \bibfield  {author} {\bibinfo {author} {\bibfnamefont {R.~P.}\ \bibnamefont
  {Kaur}}, \bibinfo {author} {\bibfnamefont {D.~F.}\ \bibnamefont {Agterberg}},
  \ and\ \bibinfo {author} {\bibfnamefont {M.}~\bibnamefont {Sigrist}},\
  }\bibfield  {title} {\enquote {\bibinfo {title} {Helical vortex phase in the
  noncentrosymmetric {CePt$_3$Si}},}\ }\href {\doibase
  10.1103/PhysRevLett.94.137002} {\bibfield  {journal} {\bibinfo  {journal}
  {Phys. Rev. Lett.}\ }\textbf {\bibinfo {volume} {94}},\ \bibinfo {pages}
  {137002} (\bibinfo {year} {2005})}\BibitemShut {NoStop}%
\bibitem [{\citenamefont {Agterberg}(2012)}]{Agterberg2012}%
  \BibitemOpen
  \bibfield  {author} {\bibinfo {author} {\bibfnamefont {D.~F.}\ \bibnamefont
  {Agterberg}},\ }\bibfield  {title} {\enquote {\bibinfo {title}
  {Magnetoelectric effects, helical phases, and {FFLO} phases},}\ }in\
  \href@noop {} {\emph {\bibinfo {booktitle} {Non-Centrosymmetric
  Superconductors}}},\ \bibinfo {editor} {edited by\ \bibinfo {editor}
  {\bibfnamefont {E.}~\bibnamefont {Bauer}}\ and\ \bibinfo {editor}
  {\bibfnamefont {M.}~\bibnamefont {Sigrist}}}\ (\bibinfo  {publisher}
  {Springer, Berlin},\ \bibinfo {year} {2012})\ pp.\ \bibinfo {pages}
  {155--170},\ \bibinfo {note} {arXiv:1106.0352}\BibitemShut {NoStop}%
\bibitem [{\citenamefont {Smidman}\ \emph {et~al.}(2017)\citenamefont
  {Smidman}, \citenamefont {Salamon}, \citenamefont {Yuan},\ and\ \citenamefont
  {Agterberg}}]{Smidman2017}%
  \BibitemOpen
  \bibfield  {author} {\bibinfo {author} {\bibfnamefont {M.}~\bibnamefont
  {Smidman}}, \bibinfo {author} {\bibfnamefont {M.~B.}\ \bibnamefont
  {Salamon}}, \bibinfo {author} {\bibfnamefont {H.~Q.}\ \bibnamefont {Yuan}}, \
  and\ \bibinfo {author} {\bibfnamefont {D.~F.}\ \bibnamefont {Agterberg}},\
  }\bibfield  {title} {\enquote {\bibinfo {title} {Superconductivity and
  spin-orbit coupling in non-centrosymmetric materials: a review},}\ }\href
  {\doibase 10.1088/1361-6633/80/3/036501} {\bibfield  {journal} {\bibinfo
  {journal} {Rep. Prog. Phys.}\ }\textbf {\bibinfo {volume} {80}},\ \bibinfo
  {pages} {036501} (\bibinfo {year} {2017})}\BibitemShut {NoStop}%
\bibitem [{\citenamefont {He}\ \emph {et~al.}(2022)\citenamefont {He},
  \citenamefont {Tanaka},\ and\ \citenamefont {Law}}]{HeLaw2022}%
  \BibitemOpen
  \bibfield  {author} {\bibinfo {author} {\bibfnamefont {James~Jun}\
  \bibnamefont {He}}, \bibinfo {author} {\bibfnamefont {Yukio}\ \bibnamefont
  {Tanaka}}, \ and\ \bibinfo {author} {\bibfnamefont {K.~T.}\ \bibnamefont
  {Law}},\ }\bibfield  {title} {\enquote {\bibinfo {title} {A phenomenological
  theory of superconductor diodes},}\ }\href {\doibase
  10.1088/1367-2630/ac6766} {\bibfield  {journal} {\bibinfo  {journal} {New J.
  Phys.}\ }\textbf {\bibinfo {volume} {24}},\ \bibinfo {pages} {053014}
  (\bibinfo {year} {2022})}\BibitemShut {NoStop}%
\bibitem [{\citenamefont {Hasan}\ \emph {et~al.}(2024)\citenamefont {Hasan},
  \citenamefont {Shaffer}, \citenamefont {Khodas},\ and\ \citenamefont
  {Levchenko}}]{Shaffer2024}%
  \BibitemOpen
  \bibfield  {author} {\bibinfo {author} {\bibfnamefont {Jaglul}\ \bibnamefont
  {Hasan}}, \bibinfo {author} {\bibfnamefont {Daniel}\ \bibnamefont {Shaffer}},
  \bibinfo {author} {\bibfnamefont {Maxim}\ \bibnamefont {Khodas}}, \ and\
  \bibinfo {author} {\bibfnamefont {Alex}\ \bibnamefont {Levchenko}},\
  }\bibfield  {title} {\enquote {\bibinfo {title} {Supercurrent diode effect in
  helical superconductors},}\ }\href {\doibase 10.1103/PhysRevB.110.024508}
  {\bibfield  {journal} {\bibinfo  {journal} {Phys. Rev. B}\ }\textbf {\bibinfo
  {volume} {110}},\ \bibinfo {pages} {024508} (\bibinfo {year}
  {2024})}\BibitemShut {NoStop}%
\bibitem [{\citenamefont {Lu}\ \emph {et~al.}(2015)\citenamefont {Lu},
  \citenamefont {Zheliuk}, \citenamefont {Leermakers}, \citenamefont {Yuan},
  \citenamefont {Zeitler}, \citenamefont {Law},\ and\ \citenamefont
  {Ye}}]{Lu2015}%
  \BibitemOpen
  \bibfield  {author} {\bibinfo {author} {\bibfnamefont {J.~M.}\ \bibnamefont
  {Lu}}, \bibinfo {author} {\bibfnamefont {O.}~\bibnamefont {Zheliuk}},
  \bibinfo {author} {\bibfnamefont {I.}~\bibnamefont {Leermakers}}, \bibinfo
  {author} {\bibfnamefont {N.~F.~Q.}\ \bibnamefont {Yuan}}, \bibinfo {author}
  {\bibfnamefont {U.}~\bibnamefont {Zeitler}}, \bibinfo {author} {\bibfnamefont
  {K.~T.}\ \bibnamefont {Law}}, \ and\ \bibinfo {author} {\bibfnamefont
  {J.~T.}\ \bibnamefont {Ye}},\ }\bibfield  {title} {\enquote {\bibinfo {title}
  {Evidence for two-dimensional {Ising} superconductivity in gated
  {MoS$_2$}},}\ }\href {\doibase 10.1126/science.aab2277} {\bibfield  {journal}
  {\bibinfo  {journal} {Science}\ }\textbf {\bibinfo {volume} {350}},\ \bibinfo
  {pages} {1353} (\bibinfo {year} {2015})}\BibitemShut {NoStop}%
\bibitem [{\citenamefont {Saito}\ \emph {et~al.}(2016)\citenamefont {Saito},
  \citenamefont {Nakamura}, \citenamefont {Bahramy}, \citenamefont {Kohama},
  \citenamefont {Ye}, \citenamefont {Kasahara}, \citenamefont {Nakagawa},
  \citenamefont {Onga}, \citenamefont {Tokunaga}, \citenamefont {Nojima},
  \citenamefont {Yanase},\ and\ \citenamefont {Iwasa}}]{Saito2016}%
  \BibitemOpen
  \bibfield  {author} {\bibinfo {author} {\bibfnamefont {Yu}~\bibnamefont
  {Saito}}, \bibinfo {author} {\bibfnamefont {Yasuharu}\ \bibnamefont
  {Nakamura}}, \bibinfo {author} {\bibfnamefont {Mohammad~S.}\ \bibnamefont
  {Bahramy}}, \bibinfo {author} {\bibfnamefont {Yoshimitsu}\ \bibnamefont
  {Kohama}}, \bibinfo {author} {\bibfnamefont {Jianting}\ \bibnamefont {Ye}},
  \bibinfo {author} {\bibfnamefont {Yuichi}\ \bibnamefont {Kasahara}}, \bibinfo
  {author} {\bibfnamefont {Yuji}\ \bibnamefont {Nakagawa}}, \bibinfo {author}
  {\bibfnamefont {Masaru}\ \bibnamefont {Onga}}, \bibinfo {author}
  {\bibfnamefont {Masashi}\ \bibnamefont {Tokunaga}}, \bibinfo {author}
  {\bibfnamefont {Tsutomu}\ \bibnamefont {Nojima}}, \bibinfo {author}
  {\bibfnamefont {Youichi}\ \bibnamefont {Yanase}}, \ and\ \bibinfo {author}
  {\bibfnamefont {Yoshihiro}\ \bibnamefont {Iwasa}},\ }\bibfield  {title}
  {\enquote {\bibinfo {title} {Superconductivity protected by spin-valley
  locking in ion-gated {MoS$_2$}},}\ }\href {\doibase 10.1038/nphys3580}
  {\bibfield  {journal} {\bibinfo  {journal} {Nat. Phys.}\ }\textbf {\bibinfo
  {volume} {12}},\ \bibinfo {pages} {144} (\bibinfo {year} {2016})}\BibitemShut
  {NoStop}%
\bibitem [{\citenamefont {Fukuyama}\ \emph {et~al.}(1971)\citenamefont
  {Fukuyama}, \citenamefont {Ebisawa},\ and\ \citenamefont
  {Tsuzuki}}]{FukuyamaEbisawaTsuzuki1971}%
  \BibitemOpen
  \bibfield  {author} {\bibinfo {author} {\bibfnamefont {Hidetoshi}\
  \bibnamefont {Fukuyama}}, \bibinfo {author} {\bibfnamefont {Hiromichi}\
  \bibnamefont {Ebisawa}}, \ and\ \bibinfo {author} {\bibfnamefont {Toshihiko}\
  \bibnamefont {Tsuzuki}},\ }\bibfield  {title} {\enquote {\bibinfo {title}
  {Fluctuation of the order parameter and {Hall} effect},}\ }\href {\doibase
  10.1143/PTP.46.1028} {\bibfield  {journal} {\bibinfo  {journal} {Prog. Theor.
  Phys.}\ }\textbf {\bibinfo {volume} {46}},\ \bibinfo {pages} {1028} (\bibinfo
  {year} {1971})}\BibitemShut {NoStop}%
\bibitem [{\citenamefont {Aronov}\ \emph {et~al.}(1995)\citenamefont {Aronov},
  \citenamefont {Hikami},\ and\ \citenamefont
  {Larkin}}]{AronovHikamiLarkin1995}%
  \BibitemOpen
  \bibfield  {author} {\bibinfo {author} {\bibfnamefont {A.~G.}\ \bibnamefont
  {Aronov}}, \bibinfo {author} {\bibfnamefont {S.}~\bibnamefont {Hikami}}, \
  and\ \bibinfo {author} {\bibfnamefont {A.~I.}\ \bibnamefont {Larkin}},\
  }\bibfield  {title} {\enquote {\bibinfo {title} {Gauge invariance and
  transport properties in superconductors above {$T_c$}},}\ }\href {\doibase
  10.1103/PhysRevB.51.3880} {\bibfield  {journal} {\bibinfo  {journal} {Phys.
  Rev. B}\ }\textbf {\bibinfo {volume} {51}},\ \bibinfo {pages} {3880}
  (\bibinfo {year} {1995})}\BibitemShut {NoStop}%
\bibitem [{\citenamefont {Li}\ and\ \citenamefont
  {Levchenko}(2020)}]{LiLevchenko2020}%
  \BibitemOpen
  \bibfield  {author} {\bibinfo {author} {\bibfnamefont {Songci}\ \bibnamefont
  {Li}}\ and\ \bibinfo {author} {\bibfnamefont {Alex}\ \bibnamefont
  {Levchenko}},\ }\bibfield  {title} {\enquote {\bibinfo {title} {Fluctuational
  anomalous {Hall} and {Nernst} effects in superconductors},}\ }\href {\doibase
  10.1016/j.aop.2020.168137} {\bibfield  {journal} {\bibinfo  {journal} {Ann.
  Phys.}\ }\textbf {\bibinfo {volume} {417}},\ \bibinfo {pages} {168137}
  (\bibinfo {year} {2020})}\BibitemShut {NoStop}%
\bibitem [{\citenamefont {Aslamazov}\ and\ \citenamefont
  {Larkin}(1968)}]{AslamazovLarkin1968}%
  \BibitemOpen
  \bibfield  {author} {\bibinfo {author} {\bibfnamefont {L.~G.}\ \bibnamefont
  {Aslamazov}}\ and\ \bibinfo {author} {\bibfnamefont {A.~I.}\ \bibnamefont
  {Larkin}},\ }\bibfield  {title} {\enquote {\bibinfo {title} {The influence of
  fluctuation pairing of electrons on the conductivity of normal metal},}\
  }\href {\doibase 10.1016/0375-9601(68)90623-3} {\bibfield  {journal}
  {\bibinfo  {journal} {Phys. Lett. A}\ }\textbf {\bibinfo {volume} {26}},\
  \bibinfo {pages} {238} (\bibinfo {year} {1968})}\BibitemShut {NoStop}%
\bibitem [{\citenamefont {Maki}(1968)}]{Maki1968}%
  \BibitemOpen
  \bibfield  {author} {\bibinfo {author} {\bibfnamefont {Kazumi}\ \bibnamefont
  {Maki}},\ }\bibfield  {title} {\enquote {\bibinfo {title} {The critical
  fluctuation of the order parameter in type-ii superconductors},}\ }\href
  {\doibase 10.1143/PTP.39.897} {\bibfield  {journal} {\bibinfo  {journal}
  {Prog. Theor. Phys.}\ }\textbf {\bibinfo {volume} {39}},\ \bibinfo {pages}
  {897} (\bibinfo {year} {1968})}\BibitemShut {NoStop}%
\bibitem [{\citenamefont {Thompson}(1970)}]{Thompson1970}%
  \BibitemOpen
  \bibfield  {author} {\bibinfo {author} {\bibfnamefont {Richard~S.}\
  \bibnamefont {Thompson}},\ }\bibfield  {title} {\enquote {\bibinfo {title}
  {Microwave, flux flow, and fluctuation resistance of dirty type-{II}
  superconductors},}\ }\href {\doibase 10.1103/PhysRevB.1.327} {\bibfield
  {journal} {\bibinfo  {journal} {Phys. Rev. B}\ }\textbf {\bibinfo {volume}
  {1}},\ \bibinfo {pages} {327} (\bibinfo {year} {1970})}\BibitemShut {NoStop}%
\bibitem [{\citenamefont {Abrahams}\ \emph {et~al.}(1970)\citenamefont
  {Abrahams}, \citenamefont {Redi},\ and\ \citenamefont {Woo}}]{Abrahams1970}%
  \BibitemOpen
  \bibfield  {author} {\bibinfo {author} {\bibfnamefont {Elihu}\ \bibnamefont
  {Abrahams}}, \bibinfo {author} {\bibfnamefont {Martha}\ \bibnamefont {Redi}},
  \ and\ \bibinfo {author} {\bibfnamefont {James W.~F.}\ \bibnamefont {Woo}},\
  }\bibfield  {title} {\enquote {\bibinfo {title} {Effect of fluctuations on
  electronic properties above the superconducting transition},}\ }\href
  {\doibase 10.1103/PhysRevB.1.208} {\bibfield  {journal} {\bibinfo  {journal}
  {Phys. Rev. B}\ }\textbf {\bibinfo {volume} {1}},\ \bibinfo {pages}
  {208--213} (\bibinfo {year} {1970})}\BibitemShut {NoStop}%
\bibitem [{\citenamefont {Larkin}\ and\ \citenamefont
  {Varlamov}(2005)}]{VarlamovLarkin}%
  \BibitemOpen
  \bibfield  {author} {\bibinfo {author} {\bibfnamefont {A.~I.}\ \bibnamefont
  {Larkin}}\ and\ \bibinfo {author} {\bibfnamefont {A.~A.}\ \bibnamefont
  {Varlamov}},\ }\href@noop {} {\emph {\bibinfo {title} {Theory of Fluctuations
  in Superconductors}}}\ (\bibinfo  {publisher} {Oxford University Press,
  Oxford},\ \bibinfo {year} {2005})\BibitemShut {NoStop}%
\bibitem [{\citenamefont {Schmid}(1969)}]{Schmid1969}%
  \BibitemOpen
  \bibfield  {author} {\bibinfo {author} {\bibfnamefont {Albert}\ \bibnamefont
  {Schmid}},\ }\bibfield  {title} {\enquote {\bibinfo {title} {Diamagnetic
  susceptibility at the transition to the superconducting state},}\ }\href
  {\doibase 10.1103/PhysRev.180.527} {\bibfield  {journal} {\bibinfo  {journal}
  {Phys. Rev.}\ }\textbf {\bibinfo {volume} {180}},\ \bibinfo {pages} {527}
  (\bibinfo {year} {1969})}\BibitemShut {NoStop}%
\bibitem [{\citenamefont {Gor'kov}\ and\ \citenamefont
  {Eliashberg}(1968)}]{GE1968}%
  \BibitemOpen
  \bibfield  {author} {\bibinfo {author} {\bibfnamefont {L.~P.}\ \bibnamefont
  {Gor'kov}}\ and\ \bibinfo {author} {\bibfnamefont {G.~M.}\ \bibnamefont
  {Eliashberg}},\ }\bibfield  {title} {\enquote {\bibinfo {title}
  {Generalization of the {Ginzburg--Landau} equations for non-stationary
  problems in the case of alloys with paramagnetic impurities},}\ }\href@noop
  {} {\bibfield  {journal} {\bibinfo  {journal} {Sov. Phys. - JETP}\ }\textbf
  {\bibinfo {volume} {27}},\ \bibinfo {pages} {328} (\bibinfo {year}
  {1968})}\BibitemShut {NoStop}%
\bibitem [{Note1()}]{Note1}%
  \BibitemOpen
  \bibinfo {note} {Gauge invariance fixes the exact paramagnetic vertex $\Gamma
  _i(\protect \bm {q}_+,\protect \bm {q}_-)$ through the Ward identity
  $k_i\Gamma _i(\protect \bm {q}_+,\protect \bm {q}_-)=\varepsilon _{\protect
  \bm {q}_+}-\varepsilon _{\protect \bm {q}_-}$ (for a polynomial dispersion
  $\Gamma _i$ is the corresponding divided difference); it is even in $\protect
  \bm {k}$ to all orders and reduces to $v_i(\protect \bm {q})$ at linear
  order, so all odd-in-$\protect \bm {k}$ dependence of Eq.~\protect \textup
  {\hbox {\mathsurround \z@ \protect \normalfont (\ignorespaces \ref
  {eq:bubble}\unskip \@@italiccorr )}} resides in the propagators. The
  diamagnetic term is even in $\protect \bm {k}$ as well, being symmetric under
  $\protect \bm {A}_{\protect \bm {k}}\protect \bm {A}_{-\protect \bm {k}}$,
  and contributes only to the reactive part of the reciprocal
  sector.}\BibitemShut {Stop}%
\bibitem [{\citenamefont {Aslamazov}\ and\ \citenamefont
  {Varlamov}(1980)}]{AslamazovVarlamov1980}%
  \BibitemOpen
  \bibfield  {author} {\bibinfo {author} {\bibfnamefont {L.~G.}\ \bibnamefont
  {Aslamazov}}\ and\ \bibinfo {author} {\bibfnamefont {A.~A.}\ \bibnamefont
  {Varlamov}},\ }\bibfield  {title} {\enquote {\bibinfo {title} {Fluctuation
  conductivity in intercalated superconductors},}\ }\href {\doibase
  10.1007/BF00115277} {\bibfield  {journal} {\bibinfo  {journal} {Journal of
  Low Temperature Physics}\ }\textbf {\bibinfo {volume} {38}},\ \bibinfo
  {pages} {223--241} (\bibinfo {year} {1980})}\BibitemShut {NoStop}%
\bibitem [{\citenamefont {Barash}\ and\ \citenamefont
  {Galaktionov}(1993)}]{Barash1993}%
  \BibitemOpen
  \bibfield  {author} {\bibinfo {author} {\bibfnamefont {Yu.~S.}\ \bibnamefont
  {Barash}}\ and\ \bibinfo {author} {\bibfnamefont {A.~V.}\ \bibnamefont
  {Galaktionov}},\ }\bibfield  {title} {\enquote {\bibinfo {title} {Frequency
  and wave-vector dependence of the fluctuation electromagnetic response near
  the superconducting transition},}\ }\href {\doibase 10.1103/PhysRevB.48.6284}
  {\bibfield  {journal} {\bibinfo  {journal} {Phys. Rev. B}\ }\textbf {\bibinfo
  {volume} {48}},\ \bibinfo {pages} {6284--6292} (\bibinfo {year}
  {1993})}\BibitemShut {NoStop}%
\bibitem [{\citenamefont {Galaktionov}(1995)}]{Galaktionov1995}%
  \BibitemOpen
  \bibfield  {author} {\bibinfo {author} {\bibfnamefont {A.~V.}\ \bibnamefont
  {Galaktionov}},\ }\bibfield  {title} {\enquote {\bibinfo {title} {Nonlocal
  fluctuational electromagnetic response of a thin film near superconducting
  transition},}\ }\href {\doibase 10.1016/0921-4534(94)02343-3} {\bibfield
  {journal} {\bibinfo  {journal} {Physica C}\ }\textbf {\bibinfo {volume}
  {241}},\ \bibinfo {pages} {118--122} (\bibinfo {year} {1995})}\BibitemShut
  {NoStop}%
\bibitem [{\citenamefont {Dolgirev}\ \emph {et~al.}(2022)\citenamefont
  {Dolgirev}, \citenamefont {Chatterjee}, \citenamefont {Esterlis},
  \citenamefont {Zibrov}, \citenamefont {Lukin}, \citenamefont {Yao},\ and\
  \citenamefont {Demler}}]{Dolgirev2022}%
  \BibitemOpen
  \bibfield  {author} {\bibinfo {author} {\bibfnamefont {Pavel~E.}\
  \bibnamefont {Dolgirev}}, \bibinfo {author} {\bibfnamefont {Shubhayu}\
  \bibnamefont {Chatterjee}}, \bibinfo {author} {\bibfnamefont {Ilya}\
  \bibnamefont {Esterlis}}, \bibinfo {author} {\bibfnamefont {Alexander~A.}\
  \bibnamefont {Zibrov}}, \bibinfo {author} {\bibfnamefont {Mikhail~D.}\
  \bibnamefont {Lukin}}, \bibinfo {author} {\bibfnamefont {Norman~Y.}\
  \bibnamefont {Yao}}, \ and\ \bibinfo {author} {\bibfnamefont {Eugene}\
  \bibnamefont {Demler}},\ }\bibfield  {title} {\enquote {\bibinfo {title}
  {Characterizing two-dimensional superconductivity via nanoscale noise
  magnetometry with single-spin qubits},}\ }\href {\doibase
  10.1103/PhysRevB.105.024507} {\bibfield  {journal} {\bibinfo  {journal}
  {Phys. Rev. B}\ }\textbf {\bibinfo {volume} {105}},\ \bibinfo {pages}
  {024507} (\bibinfo {year} {2022})}\BibitemShut {NoStop}%
\bibitem [{\citenamefont {Chatterjee}\ \emph {et~al.}(2022)\citenamefont
  {Chatterjee}, \citenamefont {Dolgirev}, \citenamefont {Esterlis},
  \citenamefont {Zibrov}, \citenamefont {Lukin}, \citenamefont {Yao},\ and\
  \citenamefont {Demler}}]{Chatterjee2022}%
  \BibitemOpen
  \bibfield  {author} {\bibinfo {author} {\bibfnamefont {Shubhayu}\
  \bibnamefont {Chatterjee}}, \bibinfo {author} {\bibfnamefont {Pavel~E.}\
  \bibnamefont {Dolgirev}}, \bibinfo {author} {\bibfnamefont {Ilya}\
  \bibnamefont {Esterlis}}, \bibinfo {author} {\bibfnamefont {Alexander~A.}\
  \bibnamefont {Zibrov}}, \bibinfo {author} {\bibfnamefont {Mikhail~D.}\
  \bibnamefont {Lukin}}, \bibinfo {author} {\bibfnamefont {Norman~Y.}\
  \bibnamefont {Yao}}, \ and\ \bibinfo {author} {\bibfnamefont {Eugene}\
  \bibnamefont {Demler}},\ }\bibfield  {title} {\enquote {\bibinfo {title}
  {Single-spin qubit magnetic spectroscopy of two-dimensional
  superconductivity},}\ }\href {\doibase 10.1103/PhysRevResearch.4.L012001}
  {\bibfield  {journal} {\bibinfo  {journal} {Phys. Rev. Res.}\ }\textbf
  {\bibinfo {volume} {4}},\ \bibinfo {pages} {L012001} (\bibinfo {year}
  {2022})}\BibitemShut {NoStop}%
\bibitem [{\citenamefont {Liu}\ \emph {et~al.}(2025)\citenamefont {Liu},
  \citenamefont {Gong}, \citenamefont {Kim}, \citenamefont {Diessel},
  \citenamefont {Xu}, \citenamefont {Rehfuss}, \citenamefont {Du},
  \citenamefont {He}, \citenamefont {Singh}, \citenamefont {Eo}, \citenamefont
  {Henriksen}, \citenamefont {Gu}, \citenamefont {Yao}, \citenamefont
  {Machado}, \citenamefont {Ran}, \citenamefont {Chatterjee},\ and\
  \citenamefont {Zu}}]{Liu2025}%
  \BibitemOpen
  \bibfield  {author} {\bibinfo {author} {\bibfnamefont {Zhongyuan}\
  \bibnamefont {Liu}}, \bibinfo {author} {\bibfnamefont {Ruotian}\ \bibnamefont
  {Gong}}, \bibinfo {author} {\bibfnamefont {Jaewon}\ \bibnamefont {Kim}},
  \bibinfo {author} {\bibfnamefont {Oriana~K.}\ \bibnamefont {Diessel}},
  \bibinfo {author} {\bibfnamefont {Qiaozhi}\ \bibnamefont {Xu}}, \bibinfo
  {author} {\bibfnamefont {Zackary}\ \bibnamefont {Rehfuss}}, \bibinfo {author}
  {\bibfnamefont {Xinyi}\ \bibnamefont {Du}}, \bibinfo {author} {\bibfnamefont
  {Guanghui}\ \bibnamefont {He}}, \bibinfo {author} {\bibfnamefont {Abhishek}\
  \bibnamefont {Singh}}, \bibinfo {author} {\bibfnamefont {Yun~Suk}\
  \bibnamefont {Eo}}, \bibinfo {author} {\bibfnamefont {Erik~A.}\ \bibnamefont
  {Henriksen}}, \bibinfo {author} {\bibfnamefont {G.~D.}\ \bibnamefont {Gu}},
  \bibinfo {author} {\bibfnamefont {Norman~Y.}\ \bibnamefont {Yao}}, \bibinfo
  {author} {\bibfnamefont {Francisco}\ \bibnamefont {Machado}}, \bibinfo
  {author} {\bibfnamefont {Sheng}\ \bibnamefont {Ran}}, \bibinfo {author}
  {\bibfnamefont {Shubhayu}\ \bibnamefont {Chatterjee}}, \ and\ \bibinfo
  {author} {\bibfnamefont {Chong}\ \bibnamefont {Zu}},\ }\href
  {https://arxiv.org/abs/2502.04439} {\enquote {\bibinfo {title} {Quantum noise
  spectroscopy of superconducting dynamics in thin film
  {Bi$_2$Sr$_2$CaCu$_2$O$_{8+\delta}$}},}\ } (\bibinfo {year} {2025}),\ \Eprint
  {http://arxiv.org/abs/2502.04439} {arXiv:2502.04439 [cond-mat.supr-con]}
  \BibitemShut {NoStop}%
\bibitem [{\citenamefont {Levchenko}\ and\ \citenamefont
  {Kamenev}(2007)}]{Levchenko2007}%
  \BibitemOpen
  \bibfield  {author} {\bibinfo {author} {\bibfnamefont {Alex}\ \bibnamefont
  {Levchenko}}\ and\ \bibinfo {author} {\bibfnamefont {Alex}\ \bibnamefont
  {Kamenev}},\ }\bibfield  {title} {\enquote {\bibinfo {title} {Keldysh
  {Ginzburg-Landau} action of fluctuating superconductors},}\ }\href {\doibase
  10.1103/PhysRevB.76.094518} {\bibfield  {journal} {\bibinfo  {journal} {Phys.
  Rev. B}\ }\textbf {\bibinfo {volume} {76}},\ \bibinfo {pages} {094518}
  (\bibinfo {year} {2007})}\BibitemShut {NoStop}%
\bibitem [{\citenamefont {Bhatnagar}\ \emph {et~al.}(1954)\citenamefont
  {Bhatnagar}, \citenamefont {Gross},\ and\ \citenamefont {Krook}}]{BGK1954}%
  \BibitemOpen
  \bibfield  {author} {\bibinfo {author} {\bibfnamefont {P.~L.}\ \bibnamefont
  {Bhatnagar}}, \bibinfo {author} {\bibfnamefont {E.~P.}\ \bibnamefont
  {Gross}}, \ and\ \bibinfo {author} {\bibfnamefont {M.}~\bibnamefont
  {Krook}},\ }\bibfield  {title} {\enquote {\bibinfo {title} {A model for
  collision processes in gases. i. small amplitude processes in charged and
  neutral one-component systems},}\ }\href {\doibase 10.1103/PhysRev.94.511}
  {\bibfield  {journal} {\bibinfo  {journal} {Phys. Rev.}\ }\textbf {\bibinfo
  {volume} {94}},\ \bibinfo {pages} {511} (\bibinfo {year} {1954})}\BibitemShut
  {NoStop}%
\bibitem [{\citenamefont {Liu}\ \emph {et~al.}(2026)\citenamefont {Liu},
  \citenamefont {de~Miranda}, \citenamefont {Shaffer},\ and\ \citenamefont
  {Levchenko}}]{Liu2026}%
  \BibitemOpen
  \bibfield  {author} {\bibinfo {author} {\bibfnamefont {Tony}\ \bibnamefont
  {Liu}}, \bibinfo {author} {\bibfnamefont {Joaquim~Telles}\ \bibnamefont
  {de~Miranda}}, \bibinfo {author} {\bibfnamefont {Daniel}\ \bibnamefont
  {Shaffer}}, \ and\ \bibinfo {author} {\bibfnamefont {Alex}\ \bibnamefont
  {Levchenko}},\ }\href {https://arxiv.org/abs/2608.05306} {\enquote {\bibinfo
  {title} {Kinetic {Lifshitz} invariants and dynamics of nonreciprocal
  fluctuations in superconductors},}\ } (\bibinfo {year} {2026}),\ \Eprint
  {http://arxiv.org/abs/2608.05306} {arXiv:2608.05306 [cond-mat.supr-con]}
  \BibitemShut {NoStop}%
\end{thebibliography}%

\end{document}